\documentclass[%
 reprint,
 superscriptaddress,
 amsmath,amssymb,
 aps,
nofootinbib,
]{revtex4-2}

\newcommand*\subtxt[1]{_{\textnormal{#1}}}
\DeclareRobustCommand\_{\ifmmode\expandafter\subtxt\else\textunderscore\fi}

\usepackage[utf8]{inputenc}

\usepackage{graphicx}
\usepackage{dcolumn}
\usepackage{bm}
\usepackage{color}
\usepackage{hyperref}

\usepackage{csquotes}
\MakeOuterQuote{"}

\newcommand{\ms}[1]{\mbox{\scriptsize #1}}

\begin{document}

\preprint{}

\title{Room-Temperature Photonic Logical Qubits via Second-Order Nonlinearities}

\author{Stefan Krastanov}
\email{stefankr@mit.edu}
\affiliation{Department of Electrical Engineering and Computer Science, Massachusetts Institute of Technology, Cambridge, MA 02139, USA}
\affiliation{John A. Paulson School of Engineering and Applied Sciences, Harvard University, Cambridge, MA 02138, USA} 

\author{Mikkel Heuck}
\affiliation{Department of Electrical Engineering and Computer Science, Massachusetts Institute of Technology, Cambridge, MA 02139, USA}

\author{Jeffrey H. Shapiro}
\affiliation{Department of Electrical Engineering and Computer Science, Massachusetts Institute of Technology, Cambridge, MA 02139, USA}

\author{Prineha Narang}
\affiliation{John A. Paulson School of Engineering and Applied Sciences, Harvard University, Cambridge, MA 02138, USA} 

\author{Dirk R. Englund}%
\affiliation{Department of Electrical Engineering and Computer Science, Massachusetts Institute of Technology, Cambridge, MA 02139, USA}

\author{Kurt Jacobs} 
\affiliation{U.S. Army Research Laboratory, Computational and Information Sciences Directorate, Adelphi, MD 20783, USA}
\affiliation{Department of Physics, University of Massachusetts at Boston, Boston, MA 02125, USA}
\affiliation{Hearne Institute for Theoretical Physics, Louisiana State University, Baton Rouge, LA 70803, USA}

\date{\today}

\begin{abstract}

Recent progress in nonlinear optical materials and microresonators has brought quantum computing with bulk optical nonlinearities into the realm of possibility. This platform is of great interest, not only because photonics is an obvious choice for quantum networks, but also as a promising route to quantum information processing at room temperature. We propose an approach for reprogrammable room-temperature photonic quantum logic that significantly simplifies the realization of various quantum circuits, and in particular, of error correction. The key element is the programmable photonic multi-mode resonator that implements reprogrammable bosonic quantum logic gates, while using only the bulk $\chi^{(2)}$ nonlinear susceptibility. We theoretically demonstrate that just two of these elements suffice for a complete, compact error-correction circuit on a bosonic code, without the need for measurement or feed-forward control. Encoding and logical operations on the code are also easily achieved with these reprogrammable quantum photonic processors. An extrapolation of current progress in nonlinear optical materials and photonic circuits indicates that such circuitry should be achievable within the next decade.  
\end{abstract}

\maketitle

\section*{\label{sec:intro}Introduction}

Any attempt to build coherent quantum hardware is met with the relentless deleterious influence of the environment. To combat it, all of today's nascent quantum computers must be cooled to cryogenic temperatures. Superconducting quantum circuits require dilution refrigerators to eliminate thermal noise~\citep{Blumoff16, OMalley16}, and ion trap processors are cooled to below 10K to reduce collisions with stray gas molecules~\citep{Brown16}. This need for cooling poses a problem for many potential applications of quantum information processing; it greatly reduces the prospects for portable devices, and significantly impacts the cost and practicality of large scale deployment as repeaters and routers for communication networks. Even optical circuits that employ single-site defects (e.g. color centers or rare-earth impurities) require cryogenic temperatures to reduce thermal line broadening~\citep{fu2009observation,jahnke2015electron,plakhotnik2015electron}. So too do  linear optics schemes that employ detectors as their sole nonlinear element (in this case to avoid the overhead incurred by inefficient detection)~\citep{jonsson2019evaluating,young2018general}.

At present there are only a few platforms that appear to have the potential for quantum processing at both room temperature and pressure~\citep{shi2010room,venkataraman2013phase,nafradi2016room,ivady2019stabilization}. We explore photonic circuits that employ bulk optical nonlinearities as their nonlinear element is a particularly promising one. Bulk nonlinear elements not only do not suffer from thermal excitation, but due to their size they are less affected by thermal broadening. Until recently the possibility of realizing quantum devices with bulk nonlinearities seemed remote, due both to the weakness of these nonlinearities and the problem of wave-packet distortion~\citep{shapiro2006single, leung2009spectral, gea2010impossibility, he2011transverse, xu2013analytic, dove2014phase}. Substantial progress in the effective strength of the material nonlinearities, the introduction of ultra-confining cavities~\citep{hu2016design, hu2018experimental, choi2017self}, and a relatively simple solution to wave-packet distortions~\citep{heuck2019controlled, heuck2019photon, li2019photon} have changed that outlook. 


Achieving the physical technology to implement nonlinear photonic quantum circuits is not the only challenge to realizing room-temperature quantum logic. For practicality one must implement this logic using the strongest available nonlinearity, the leading-order $\chi^{(2)}$ nonlinear susceptibility, and for efficient room-temperature operation the logic and error-correction circuits should avoid  measurements or feed-forward control. Two basic approaches to information processing with photons are possible. The first is the use of single or dual-rail encoding in which each mode contains no more than one photon~\citep{Kok07}. While this has the advantage that all circuit constructions from the well-developed qubit model can be employed, this leads to complex circuits even for correcting the loss of a single photon. The smallest code for this purpose uses five modes (ten for dual-rail encoding)~\citep{Laflamme96, divincenzo1996fault}. While there is little work on minimal circuits for correcting the five-qubit code, from circuits for the seven-qubit Steane code\footnote{Note that the circuits for the seven-qubit Steane code devised in~\citep{Crow2016, Premakumar20} are written in terms of generalized CNOT gates that have four control qubits. To write these in terms of two-qubit CNOT gates requires additional ancilla qubits and CNOT gates.} we estimate that it requires a least nine additional modes and more than 30 CNOT gates. The alternative is to use bosonic codes that employ multiple photons per mode, but in this case it is far from obvious what gates and circuits are required to implement the error-correction, let alone how to realize these gates with a $\chi^{(2)}$ interaction. While explicit error correction procedures for bosonic codes have been elucidated~\citep{niu2018qudit, niu2018hardware, Michael16, Albert18, bergmann2016quantum} they all involve non-demolition or photon-number-resolving measurements. It is not yet known how to construct the unitary multi-photon operations required to replace such measurement using only a $\chi^{(2)}$ nonlinearity, or the complexity of doing so. The only unitary circuit that has been explicitly constructed to-date to correct a bosonic code is in the form of a forty-layer neural network using an idealized $\chi^{(3)}$ medium~\citep{steinbrecher2019quantum}. 

Here we propose an approach for implementing all-unitary, and thus room-temperature, quantum logic on multi-mode multi-photon states using only a fixed $\chi^{(2)}$ nonlinearity. This paradigm, which employs as its basic module a single triply-resonant cavity with a time-dependent drive, significantly reduces the complexity of the physical circuits required to implement multi-photon quantum logic in general, and error-correction in particular. The joint operation performed on the three modes by the module is controlled by the time-dependent drive. In this way the module is able to perform a wide range of three-mode multi-photon gates. We demonstrate the power of this approach by explicitly constructing a measurement-free error-correcting circuit for a two-mode bosonic code. This circuit requires just two of our three-mode modules, along with some controllable linear elements.
Our compact unitary circuits do not employ any measurements or feed-forward control, which makes them particularly useful for fast quantum routers and repeaters. However, measurements will certainly be required to read out a message or the results of a computation. Fortunately, it is straightforward to use unitary circuits in general, and our processor in particular, to enable high fidelity measurements at room temperature, even when only inefficient detectors are available. To do so it is enough to use a unitary circuit to map a single photon to a sufficiently large number of photons that can then be detected. This amplification can be implemented rapidly using a doubling process. First a $\chi^{(2)}$ nonlinearity is used to convert one photon in one mode 1 to two photons in a second mode via down-conversion. Second, a frequency conversion process (which employs a $\chi^{(2)}$ and a classical pump) is then used to transfer each of the two photons back to the first mode. Repeating this photon-doubling cycle provides exponentially fast amplification. Since measurements are only to be used at the end of a computation the additional overhead for amplification remains small. Thus, while we do not analyze this measurement method in detail here, it is clear that the lack of efficient photon detectors is not an obstacle to room-temperature quantum information processing.

In the next section we describe the control Hamiltonian realized by the driven triply-resonant cavity that forms our basic processing module and give examples of important gates that can be implemented by the module. Then we show how a full error correction process can be built from a small number of these multi-photon gates. Lastly, we discuss the materials science and fabrication challenges that must be addressed in order to realize our loss-correction circuit. By extrapolating the rate of progress in these areas over the last decade we estimate a timeline for demonstrating this circuit.

\begin{figure}
    \centering
    \includegraphics[width=\columnwidth]{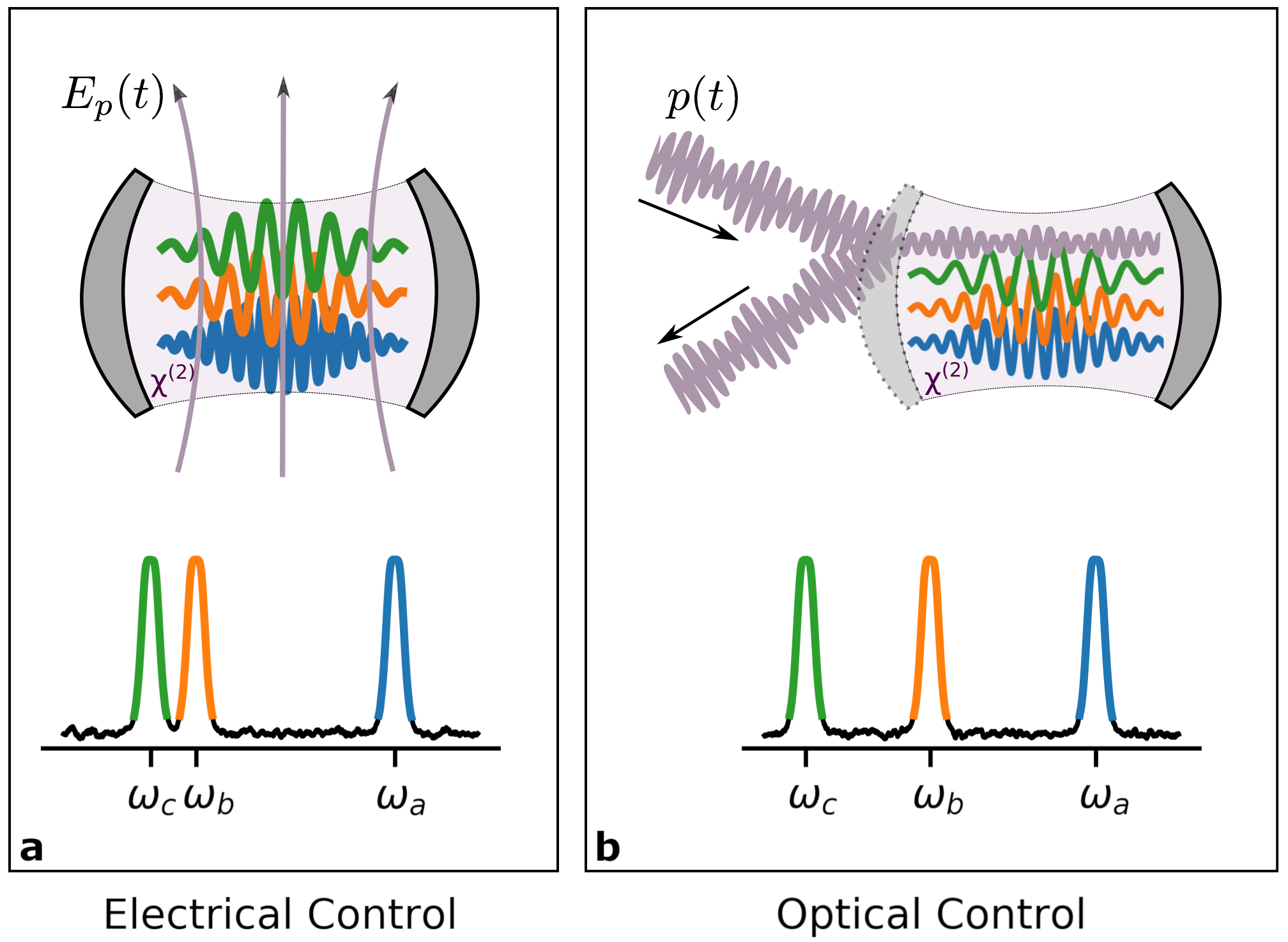}
    \caption{\emph{The triply-resonant nonlinear cavity.} The $\chi^{(2)}$ medium enables the joint control of three modes. We denote the mode operators respectively by $\hat{a}$, $\hat{b}$, and $\hat{c}$. The $\chi^{(2)}$ medium enables frequency doubling from $\hat{b}$ to $\hat{a}$, and a three-way interaction between modes $\hat{b}$, $\hat{c}$, and the control field. The control field is either (a) a classical microwave drive, $E\_p(t)$, or (b) a classical optical drive of envelope $p(t)$. This three-way interaction is effectively a linear interaction between modes $\hat{b}$ and $\hat{c}$ that is controlled by the classical drive. The combination of the fixed frequency-doubling interaction and the controlled linear interaction allows extensive control of the joint nonlinear evolution. This evolution conserves the quantity $2n\_a + n\_b + n\_c$, in which $n\_a$, $n\_b$, and $n\_c$ are the occupation numbers of the respective modes. We also depict the relative values of the frequencies of the three modes; in (a) the frequencies of modes $\hat{b}$ and $\hat{c}$ are separated by the much smaller frequency of the microwave drive.
    }
    \label{fig:cartoon-cavity}
\end{figure}

\section*{Results}

\subsection*{\label{sec:hamiltonian-and-unitaries}A Controllable Three-Mode Cavity}


We consider three resonant modes of a cavity in a $\chi^{(2)}$ medium, with respective mode operators $\hat{a}$, $\hat{b}$, and $\hat{c}$, and frequencies $\omega\_a$, $\omega\_b$, and $\omega\_c$. We neglect any dissipative dynamics until later sections where we discuss hardware implementations. The cavity is driven by a coherent classical pump with frequency $\omega\_p$. We depict it in Fig.~\ref{fig:cartoon-cavity}, in which the pump may be a microwave frequency electric field or an optical drive. By choosing the frequencies to satisfy 
\begin{align}
    \omega\_a = 2\omega\_b \\
    \omega\_c + \omega\_p =\omega\_b, 
\end{align}
the $\chi^{(2)}$ medium couples the modes via the Hamiltonian 
\begin{align}
      \hat{H}_{\ms{nl}}(t) = \hbar \left[ \chi \hat{a}\hat{b}^{\dagger 2} + g(t) \hat{b}^\dagger\hat{c}+ \mbox{H.c.} \right], 
\end{align}
in which $\chi$ is the coupling rate of the $\chi^{(2)}$ nonlinearity, $g(t)$ is the coherent amplitude of the classical pump, and we have moved to the rotating frame of the oscillators. Since Schr\"{o}dinger's equation contains $H/\hbar$, it is the rates $\chi$ and $g(t)$ that determine the dynamics. If we measure time in units of $1/\chi$ then all rates are divided by $\chi$, and the dynamics is determined by 
\begin{align}
      \frac{\hat{H}_{\ms{nl}}(t)}{\hbar\chi} =  \hat{a}\hat{b}^{\dagger 2} + p(t) \hat{b}^\dagger\hat{c}+ \mbox{H.c.}, 
\end{align}
in which we have defined $p(t) = g(t)/\chi$. Thus up to a scaling of time, the dynamics is entirely determined by the rate parameter $p(t)$. We will not have to introduce any additional rate parameters (and thus any additional timescales) until we consider loss in the section on hardware considerations. In that section we will express our rate parameters in terms of the physical parameters of realistic devices. 

Note that the second term in $\hat{H}_{\ms{nl}}(t)$, which is controlled via the amplitude of the pump, is merely a linear coupling between modes $\hat{b}$ and $\hat{c}$. This interaction cannot by itself generate a universal set of quantum gates~\citep{weedbrook2012gaussian, bartlett2002efficient}. It turns out, however, that it can do so when combined with the time-independent frequency doubling interaction. 

We denote the number of photons in the three modes respectively by $n\_a$, $n\_b$, and $n\_c$, and the corresponding operators for the photon number by $\hat{n}_a$, $\hat{n}_b$, and $\hat{n}_c$. Since the Hamiltonian commutes with $2\hat{n}_a + \hat{n}_b + \hat{n}_c$, the value of that observable is preserved. The Hamiltonian cannot, therefore, mix subspaces defined by different integer values of $2n\_a + n\_b + n\_c$. Nevertheless, it does provide complete control within each subspace by virtue of the fact that the repeated commutators of $\hat{a}\hat{b}^{\dagger 2}$ and $\hat{b}^\dagger\hat{c}$ generate a complete Lie algebra for all such subspaces~\citep{lloyd1999quantum,jacobs2007engineering,krastanov2015universal,niu2018qudit}. It is this fact that provides the power of our processing unit. 

In general, to implement quantum gates between the three modes we will need to generate a set of distinct evolutions, one for each of the $2\hat{n}_a + \hat{n}_b + \hat{n}_c=\mathrm{const}$ subspaces. We can do that with a single control pulse, as for each subspace, there are many choices for $p(t)$ that generate the same unitary operation. We can use numerical search methods to find a control function $p(t)$ that simultaneously generates the required evolution for each of the set of subspaces. Naturally we wish to find the control that implements a given gate in the shortest time, a challenge solved as described below.

Lastly, the modes of this quantum processor will need to be actively coupled to the waveguides that carry the quantum states to be processed. Otherwise, the process of capturing the content of the waveguides will be too slow due to the necessarily-high quality factor of the modes of the processor. To actively couple the cavity modes to the waveguides we envision using the method given in~\citep{heuck2019photon}.

\subsection*{\label{ssec:unitaries}Compiling Unitary Operations}

\begin{figure*}
    \centering
    \includegraphics[width=\textwidth]{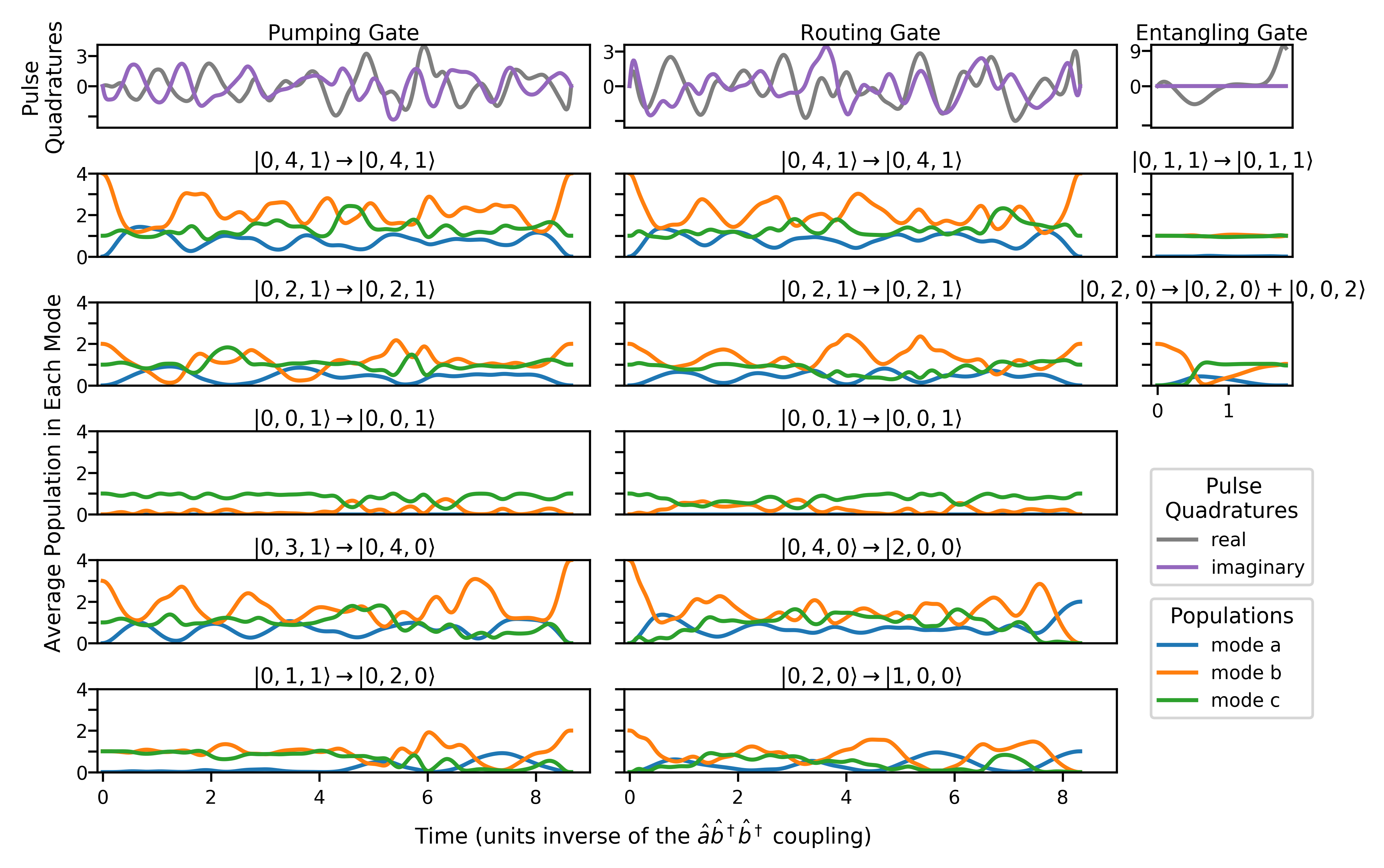}
    \caption{\emph{The control pulses implementing the three gates that are used to build our error-correction circuit.} The top row shows the real and imaginary parts of the control pulses for each gate. The following rows show how the populations of the modes evolve under each gate for a given initial state. The optimizer produces pulses $p(t)$ such that each of the desired transformations leads to constructive interference at the exact same time. Shorter pulses are possible, at the expense of higher power and bandwidth requirements~\citep{methods_inprep}, up to a point at which the pulse is too short to perform even a single complete oscillation in a subspace defined by an integer value of $2n\_a+n\_b+n\_c$.}
    \label{fig:compiled-gates}
\end{figure*}

To find the control pulse $p(t)$ required to implement a given unitary operation we employ numerical search methods, an approach often referred to as optimal control~\citep{khaneja2005optimal,heeres2017implementing,methods_inprep}. We introduce a parameterization for $p(t)$ as a piece-wise constant signal in which the duration of each interval is variable. This parameterization is essential because the always-on frequency-doubling component of the Hamiltonian necessitates optimizing the length of the pulse. In order to avoid unphysical pulses, we constrain both the duration and amplitude of each interval by the use of sigmoid functions. The full expression for the resulting unitary operation is 
\begin{align}
\hat{U}(\mathrm{\mathbf{v}}) & = 
\prod_{l=1}^{s} \exp\left\{
-i \left[ \operatorname{f}(X_l,P_l)\hat{b}^\dagger\hat{c}
+\sigma(T_l)\hat{a}\hat{b}^{\dagger 2} +\mbox{H.c.}
\right] \right\}  
\end{align} 
where 
\begin{align} 
 \operatorname{f}(X_l,P_l) & =\arctan(X_l)+i\arctan(P_l) \\
\sigma(T_l) & = \frac{\Delta \tau}{1 + \exp(-T_l)}
\end{align}
and $\mathbf{v} = \{ X_l, P_l, T_l : l = 1,\ldots, N \}$ is the set of parameters that defines the pulse. 
The parameters $\{ X_l : X_l \in \mathbb{R} \}$ and $\{ P_l : P_l \in \mathbb{R} \}$ are related to the quadrature of the pulse, which is constrained to the interval $[-1,1]$ by $\arctan$, while the $\{ T_l : T_l \in \mathbb{R} \}$ are related to the duration of each segment, which is constrained to the interval $[0,\Delta \tau]$. We fix the number of piecewise-constant intervals, $s$, as well as the relative unitless time scale $\Delta \tau$. 

Consistently good performance is obtained even with $s<60$. This permits the use of standard automatic differentiation tools, without the need for approximations such as GRAPE~\citep{khaneja2005optimal}. Our parameterization also has the advantage that it does not allow for pathological pulses. Once we have obtained a piecewise constant control function for a given gate, we use GRAPE and standard regularization techniques to smooth out the pulse, ensuring it has both reasonable bandwidth and power. Throughout the optimization, the robustness of the control to calibration errors is verified. The time scale $\Delta \tau$ is shortened until a threshhold is reached at which the control pulse is no longer robust. The above approach to generating control functions, together with a number of symbolic optimizations, will be presented in detail in~\citep{methods_inprep}.

\subsection*{Examples of Programmable Gates} 

Through the use of our implicitly constrained optimal control method, we can perform with high fidelity any gate that keeps $2n\_a+n\_b+n\_c$ constant. If the length of the control pulse is unconstrained, and dissipation is neglected, we can achieve fidelities arbitrary close to unity. For gates reported here, we constrain the duration of the control pulses as much as possible before reaching unitary fidelities lower than $0.999$. In later sections considering hardware implementations, we also describe the effects of dissipation. Here we describe a number of important unitary operations that fulfill that constraint, some of which are also depicted in Fig.~\ref{fig:compiled-gates}. More general unitary operations can be performed by reshuffling the modes of the three-mode processors, as seen in later sections. Given the long cavity lifetimes requires for these operations, reshuffling necessitates rapid catch and release of photons from and into the connected waveguides, e.g., by using active control as done in \citep{heuck2019photon}.

Throughout the following paragraphs we will use the notation $|n\_a n\_b n\_c\rangle$ to denote a Fock state with $n\_a$, $n\_b$, and $n\_c$ photons in modes $\hat{a}$, $\hat{b}$, and $\hat{c}$, respectively.


We begin with the Toffoli Gate, which is a three-qubit non-Clifford gate, distinguished by the fact that together with just the single-qubit Hadamard gate it enables universal quantum circuits~\citep{shi2002both,aharonov2003simple}. Of particular relevance for our purposes is the fact that it usually requires six two-qubit CNOT gates to implement~\citep{barenco1995elementary,shende2008cnot}, while our realization requires only a single application of the three-mode processor. We realize the gate in the Hadamard basis (i.e., our gate is a Phase gate with two control qubits) for photonic qubits encoded in a single- or dual-rail configuration. In this basis the Toffoli unitary maps all joint Fock states to themselves except for the state $|111\rangle$ to which it applies a $\pi$ phase.


We also define a conditional routing gate as one that swaps the state of two modes depending on the state of a third mode. This class of gates is useful for breaking down conditional multi-qudit operations into smaller units. We first route the target mode to a particular waveguide, based on the state of the control mode, and we perform the appropriate single-mode quantum operation in the new physical location of the target mode. Such routing is indispensable, if our goal is to avoid measurements in error correcting circuits, as measurements usually require hardware at cryogenic temperatures. Typically, a non-demolition measurement is performed by entangling the required information with an ancilla and performing a demolition measurement on the ancilla. The result, a classical bit, is then fed forward through a classical computer that decides what quantum operation to perform next. We avoid the measurement and classical decisions through coherent quantum feedback~\citep{lloyd2000coherent,Jacobs2014coherent}, where we simply perform a multi-mode quantum gate conditioned on the ancilla. The realization for the routing gate suggested below is what we use in our bosonic error-correcting circuit, but other setups are feasible as well. Below $|n\_a n\_b n\_c\rangle$ denotes a Fock state with $n\_a$, $n\_b$, and $n\_c$ photons in modes $\hat{a}$, $\hat{b}$, and $\hat{c}$, respectively. The $\hat{c}$ mode is the control, the $\hat{b}$ mode is the input, and $\hat{a}$ and $\hat{b}$ are the possible outputs:

\begin{align}
|040\rangle & \mapsto |200\rangle \\
|020\rangle & \mapsto |100\rangle \\
|041\rangle & \mapsto |041\rangle \\
|021\rangle & \mapsto |021\rangle \\
|001\rangle & \mapsto |001\rangle
\end{align}

When used in the error correcting circuits described in later sections, mode $\hat{c}$ will contain an ancillary photon on which routing will be conditioned, while mode $\hat{b}$ will contain one of the two modes of our multimode bosonic code. A second processor will be used for other modes.


Focusing further on the error-correcting functionality, we need a gate that can correct for photon loss in a codeword. For the code we employ we require the gate to preserve the states $|001\rangle$, $|021\rangle$, and $|041\rangle$, and accomplish the mapping 
\begin{align}
|031\rangle & \mapsto |040\rangle ,\\
|011\rangle & \mapsto |020\rangle . 
\end{align}

This operation is necessary for reverting photon loss in the code mode stored in $\hat{b}$, while storing information about the occurrence of that loss in mode $\hat{c}$. Again, we will need two three-mode processors, each acting on one of the modes making up our error correcting code. Each of the physical modes of the code will be stored in the corresponding $\hat{b}$ oscillators.


To complete our error-correction circuit we use a gate that entangles two modes. We require this operation because one of the code words is an entangled state, and the loss of a photon breaks this entanglement. This gate provides the mapping 
\begin{align}
|011\rangle & \mapsto |011\rangle \\
|020\rangle & \mapsto \frac{|020\rangle+|002\rangle}{\sqrt{2}} 
\end{align}
This gate is also a symmetrizing operation for the state of the modes $\hat{b}$ and $\hat{c}$. It is these two oscillators $\hat{b}$ and $\hat{c}$ that will contain the two modes of our error correcting code.


The above gates are only a few of the many operations that the triply-resonant cavity processor can perform. Among these gates are those important for the processing of unprotected single photon-states, and operations that enable unitary modification and number-resolved measurements on modes with more than one photons, including bosonic codes. Importantly, these operations are performed with a single use of the triply-resonant cavity, while otherwise they would require complete circuits with multiple discrete operations. This leads to drastically simpler overall circuits, at the expense of requiring this more sophisticated and difficult to fabricate triply-resonant optical resonator.

\begin{figure}
    \centering
    \includegraphics[width=\columnwidth]{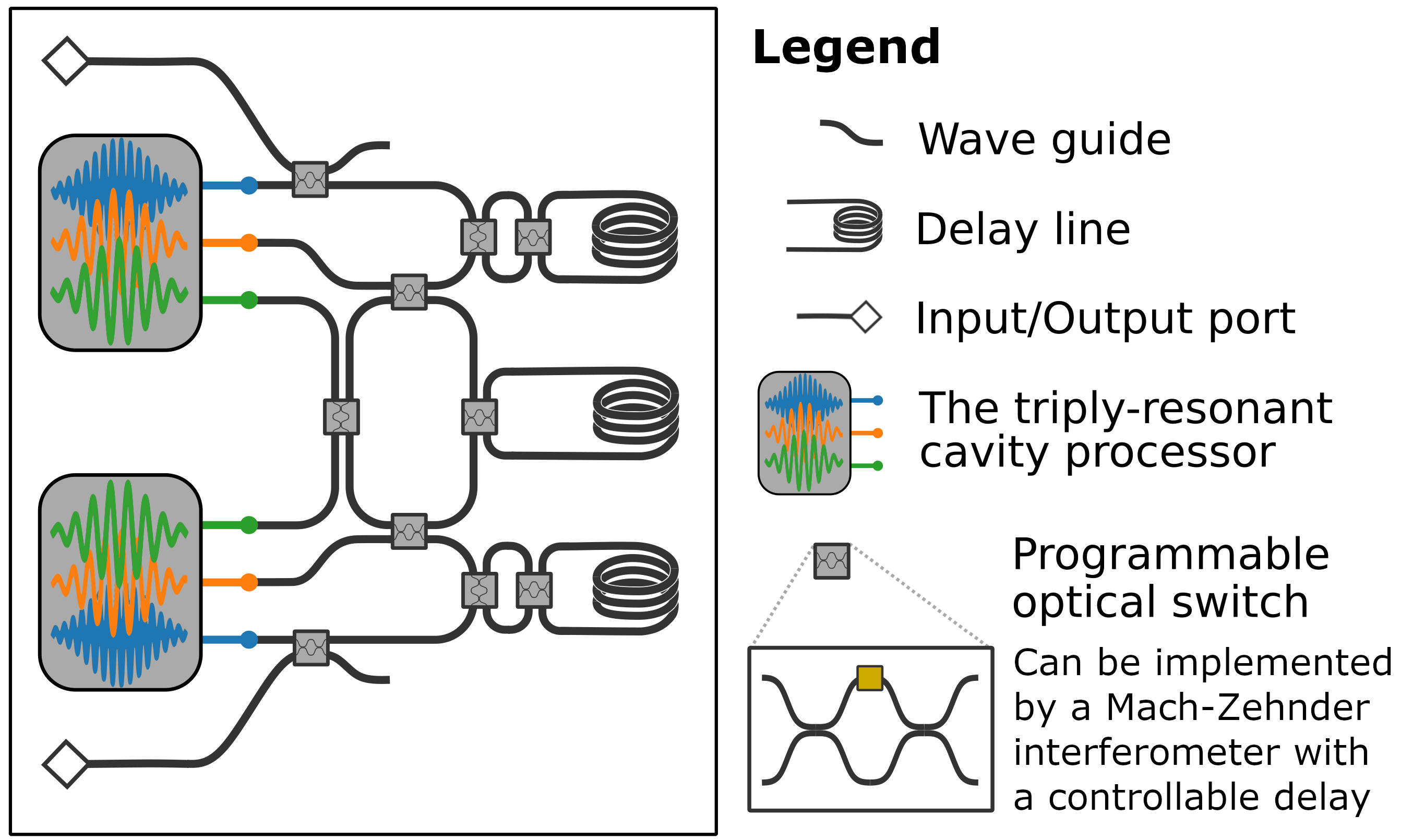}
    \caption{\emph{Our minimal architecture for error correction of bosonic codes, readily expandable to larger tasks.} The circuit depicted can be used to correct a single-photon loss using a two-mode bosonic code. The circuit consists of two cavity processors, which for the most part process each mode of the code separately, and a small network of reprogrammable beam splitters and delay lines. These are used to reroute states between the modes of processors as necessary. Each cavity processor is also capable of performing many multi-qubit gates for single- and dual-rail encoded qubits, as well as preparing and manipulating higher-number Fock states. The network of programmable beam splitters between the processors and the delay lines can also be expanded to a fully connected network, enabling universal rerouting between the three modes of each processor for general-purpose quantum computation. The programmable beam splitters can be implemented as Mach-Zehnder interferometers (as shown in the inset) with two 50/50 beam splitters and a programmable delay (the orange medium in the diagram). Classical electronics will be necessary to ensure the pacing of various operations in this device, but no feedback or decision circuitry is necessary, as the approach is measurement free.}
    \label{fig:pingpong}
\end{figure}

\begin{figure*}
    \centering
    \includegraphics[width=\textwidth]{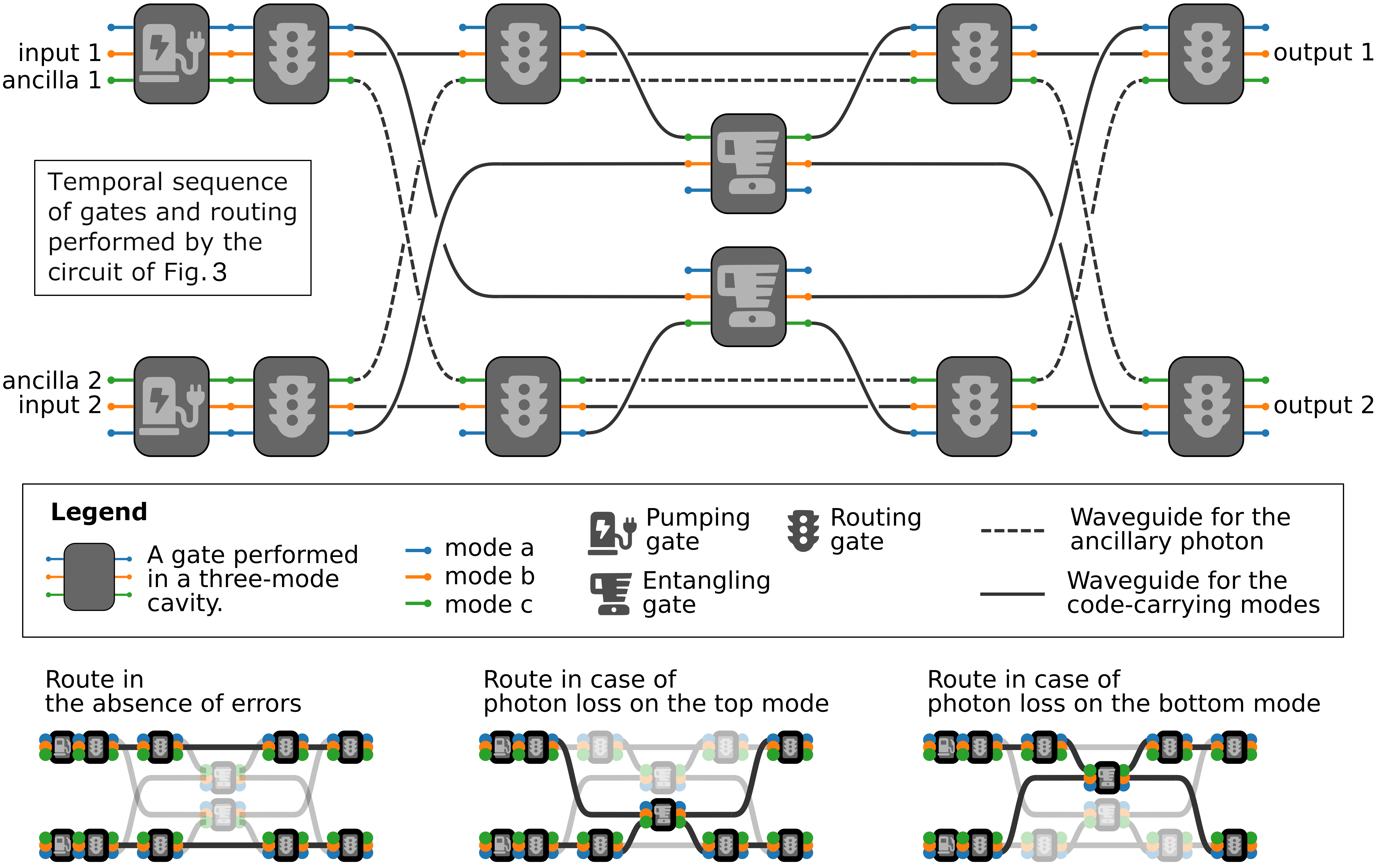}
    \caption{\emph{The error-correcting circuit, unrolled in time.} The horizontal axis represents the flow of time, depicting how a pair of triply-resonant cavities is being used. This circuit would be executed on the hardware depicted in Fig.\ \ref{fig:pingpong}. The main drawing is the sequence of operations that we need to perform in parallel in the two triply-resonant cavities in order to perform the error correction. After placing the code and ancilla modes in the appropriate cavity modes, we accomplish the initial pumping and routing gates. After that, we need to shuffle the ancillary modes by releasing them in the appropriate waveguides. The spatial modes into which the code states are moved depend on the state of the ancillas, thanks to the conditional routing gates. As the ancillas contain information about the presence of photon-loss errors, this lets us perform operations conditioned on the loss of a photon, by performing the two conditional branches in parallel in different physical locations of the circuit. The conditional routing gates then act in reverse, ensuring that all spatial modes end in the same location, without breaking the bijectivity required for any quantum circuit. The various spatial modes employed can be seen in the bottom insets of the figure. Supplementary Figure 1 provides a more detailed rendition. Importantly, as seen in Fig.\ \ref{fig:pingpong}, we do not need 12 triply-resonant cavities as depicted above, rather only 2 cavities with a network of waveguides and programmable beam splitters~\citep{carolan2015universal} that can route the spatial modes as necessary, so that each cavity can be used repeatedly. The gate pictographs are taken from the Font Awesome icon set.}
    \label{fig:ecc-circuit}
\end{figure*}

\subsection*{Measurement-free Error Correction}


We demonstrate the versatility of our control protocol by constructing an error-correcting circuit around the three-mode processor. The circuit we obtain is not only simple and short, but it also does not require any measurement operations or classical feed-forward control.

We choose the following code, encoding a single qubit in two separate (spatial) bosonic modes, whose logical states are given by \begin{align}
    |1\rangle_{\ms{L}} & = |22\rangle , \\ 
    |0\rangle_{\ms{L}} & = \frac{|40\rangle+|04\rangle}{\sqrt2} . 
\end{align}
This two-mode code allows correction for the loss of a single photon from either mode. For a channel that has a 10\% probability of a single photon loss for each mode this implies an 81\% chance of transmission without error, 18\% chance of transmission with a correctable error, and a 1\% chance of transmission with an uncorrectable error. We choose this code because it is possible to perform the correction process with operations that conserve the quantity $2n\_a+n\_b+n\_c$, so long as one is judicious in choosing these operations.

We must first consider the effect of a photon loss on the code. The loss of a photon on the first mode is described by the action of $\hat{a}\hat{I}$. This transforms the initial code state $|C\rangle = \alpha |0\rangle_{\ms{L}} + \beta |1\rangle_{\ms{L}}$ into the error state  $|E\_1\rangle=\alpha|12\rangle+\beta|30\rangle$. Similarly the loss of a photon from the second mode produces the error state $|E\_2\rangle=\alpha|21\rangle+\beta|03\rangle$. For each of these two errors we need to perform a different correction procedure. Typically this is achieved by a non-demolition measurement that projects the state of the system onto either the logical subspace or one of the error subspaces, followed by a unitary correction operation conditioned on the measurement result. We sidestep these requirements by using coherent control. We employ two quantum ancillas, initialized to contain single photons, on which routing gates will be conditioned. Thus, our correction procedure involves the following steps. First, we put the information about the presence of an error in the ancillas by using two conditional pumping gates acting in parallel (the code modes are each placed in a $\hat{b}$ mode, while the ancillary photons are in the corresponding $\hat{c}$ modes), resulting in the following transformation of the overall ancillas-code state:
\begin{align}
|11\rangle \otimes |C  \rangle ~ & 
  \mapsto
  |11\rangle \otimes |C  \rangle \\
|11\rangle \otimes |E\_1\rangle &
  \mapsto
  |01\rangle \otimes |F_1\rangle  \\
|11\rangle \otimes |E\_2\rangle &
  \mapsto
  |10\rangle \otimes |F_2\rangle,
\end{align}
where $|F_1\rangle=\alpha|22\rangle+\beta|40\rangle$ and $|F_2\rangle=\alpha|22\rangle+\beta|04\rangle$. The feed-forward solution would have measured the ancillas and performed different operations depending on the measurement, but as already mentioned that would be slow and require additional cooled hardware and classical decision circuitry. Instead, we perform the following unitary operation (as before, the left multiplier in the tensor product $|c_1 c_2\rangle$ denotes the content of the ancillary $\hat{c}$ modes of each of the two processors, and $|C  \rangle$ denotes the two modes of the error correcting code, stored in the $\hat{b}$ oscillators of the two processor):
\begin{align}
|11\rangle \otimes |C  \rangle ~ & 
  \mapsto
  |11\rangle \otimes |C\rangle \\
|01\rangle \otimes |F_1\rangle &
  \mapsto
  |01\rangle \otimes |C\rangle  \\
|10\rangle \otimes |F_2\rangle &
  \mapsto
  |10\rangle \otimes |C\rangle,
\end{align}
Without the ancillas, this operation would be impossible as it would break the bijectivity of the unitary operator by mapping many states to one. The conditional routing gates are crucial for the performance of this operation -- depending on the ancillas, they route the modes containing the code to different spatial modes that perform $|F_1\rangle\mapsto|C\rangle$ and $|F_2\rangle\mapsto|C\rangle$ independently and in parallel. The conditional routing gates then ensure that all three paths end up in the same spatial modes at the end of the circuit. The error-correcting circuit can be seen in Fig.~\ref{fig:pingpong} as a suggested physical layout, and in Fig.~\ref{fig:ecc-circuit} as a sequence of abstract gates.

\subsection*{Encoding Operation} Encoding a qubit in the two-mode code is particularly simple using the three-mode processor. To do so we have to perform the operation 

\begin{align}
    |00\rangle & \mapsto |22\rangle , \\
    |10\rangle & \mapsto \frac{|40\rangle+|04\rangle}{\sqrt2} .
\end{align}
Given that we already have access to the entangling gate, encoding can be done by putting the unprotected photonic qubit in cavity $\hat{c}$ and putting ancilla photons in cavities $\hat{a}$ and $\hat{b}$. Then we perform the partial encoding gate
\begin{align}
|111\rangle & \mapsto |200\rangle \\
|110\rangle & \mapsto |110\rangle,
\end{align}
thus mapping the state $\alpha |1\rangle + \beta |0\rangle$ in $\hat{c}$ to the precursor of the two-mode code $\alpha |11\rangle + \beta |20\rangle$ in $\hat{a}$ and $\hat{b}$. As before, here $|n\_a n\_b n\_c\rangle$ denotes a Fock state with $n\_a$, $n\_b$, and $n\_c$ photons the modes $\hat{a}$, $\hat{b}$, and $\hat{c}$ of a single processor. Turning this into the complete code state requires a simple application of the entangling operation already discussed above. These two operations preserve the constant of motion $2\hat{n}_a + \hat{n}_b + \hat{n}_c$ and as such can be compiled to a single control pulse performed in a single triply-resonant cavity.

\subsection*{Two-Qubit Logical Operations}

Single-qubit rotations in the logical space of the two-mode code can be realized by using our three-mode processors, as such rotations do preserve $2n\_a+n\_b+n\_c$. Moreover, two qubit logical operations can also be performed. For instance, consider a CPHASE gate, which together with the single-qubit rotations form a universal set. We need to perform the operation
\begin{equation} 
    |22\rangle\otimes|22\rangle \mapsto -|22\rangle\otimes|22\rangle,
\end{equation}
while mapping all other combinations of $\left\{|22\rangle, \frac{|40\rangle+|04\rangle}{\sqrt2}\right\}^{\otimes2}$ to themselves. Our three-mode processor can perform this operation by acting on just two of the four modes making up the two logical qubits. If we index each mode as, e.g., $|2_i2_j\rangle\otimes|2_k2_l\rangle$, we need to act only on modes $j$ and $k$, by first transferring them to modes $b$ and $c$ of the cavity and designing a pulse to perform the operation
\begin{equation} 
    |022\rangle \mapsto -|022\rangle,
\end{equation}
while mapping all of $|000\rangle$, $|002\rangle$, $|020\rangle$, $|004\rangle$, $|040\rangle$, $|024\rangle$, $|042\rangle$, and $|044\rangle$ to themselves.
The overall CPHASE operation on the four physical modes forming the two logical qubits takes the form,
\begin{alignat}{8}
    |2\underline{2}\rangle
    &\otimes &&
    |\underline{2}2\rangle &
    \,\mapsto \, &&
    -|2\underline{2}\rangle
    &\otimes &&
    |\underline{2}2\rangle,
    \\
    \frac{|4\underline{0}\rangle+|0\underline{4}\rangle}{\sqrt 2}
    &\otimes &&
    |\underline{2}2\rangle &
    \,\mapsto \, &&
    \frac{|4\underline{0}\rangle+|0\underline{4}\rangle}{\sqrt 2}
    &\otimes &&
    |\underline{2}2\rangle,
    \\
    |2\underline{2}\rangle
    &\otimes &&
    \frac{|\underline{4}0\rangle+|\underline{0}4\rangle}{\sqrt 2} &
    \,\mapsto \, &&
    |2\underline{2}\rangle
    &\otimes &&
    \frac{|\underline{4}0\rangle+|\underline{0}4\rangle}{\sqrt 2},
    \\
    \frac{|4\underline{0}\rangle+|0\underline{4}\rangle}{\sqrt 2}
    &\otimes &&
    \frac{|\underline{4}0\rangle+|\underline{0}4\rangle}{\sqrt 2} &
    \,\mapsto \, &&
    \frac{|4\underline{0}\rangle+|0\underline{4}\rangle}{\sqrt 2}
    &\otimes &&
    \frac{|\underline{4}0\rangle+|\underline{0}4\rangle}{\sqrt 2},
\end{alignat}
where the underlined modes are the ones that are manipulated inside of a three-mode processor. Notice that a phase is gained only in the first row, where both of the modes in the processor (the underlined modes) have two photons.
Such an operation can be performed directly by our processors or, if shorter and simpler control pulses are desired, by first using the $\chi^{(2)}$ interaction to upconvert them to lower photon numbers.

\subsection*{Comparison with Other Approaches}
\label{subsec_other}

Comparisons with other codes and types of hardware require care because the various systems have significant differences. Nevertheless, we elucidate how our control protocol substantially reduces the depth of a typical circuit and removes the need for entire classes of expensive operations. As discussed in the introduction, error-correction procedures have been proposed for bosonic codes, but these require non-demolition or photon-number-resolving measurements, and it has not yet been described how such measurements can be replaced by unitary operations generated by a $\chi^{(2)}$ nonlinearity. We can however, compare our circuit to the explicit correction circuit presented in~\citep{steinbrecher2019quantum}. 

One way to compare the efficiency of circuits is to examine how long each takes relative to the characteristic unit of time for the given hardware. The circuit we have constructed above requires six gates, for a total of forty units of time (relative to the $\chi^{(2)}$ coupling strength) and 4 transfers in and out of cavities. The correction circuit employing the quantum optical neural network (QONN) architecture~\citep{steinbrecher2019quantum}, which is the closest analog of our hardware, requires 40 layers, resulting also in 40 units of time, but since it uses a $\chi^{(3)}$ rather than a $\chi^{(2)}$ medium, the nonlinearity is significantly weaker, so that the circuit takes longer in real time. Furthermore, the QONN circuit requires 40 transfers in and out of the nonlinear-cavities (one for each layer), ten times more than our architecture. 

One can instead implement photonic quantum logic by using only the vacuum and 1-photon Fock states to encode qubits (i.e., a single- or dual-rail encoding). The smallest error-correcting code in this setting requires five physical qubits~\citep{Laflamme96}. The logic required to determine the error syndrome for this code requires sixteen CNOT gates and four auxiliary qubits~\citep{divincenzo1996fault}. The auxiliary qubits can either be measured, in which case the error can be determined using a classical computer, or a unitary circuit could process the auxiliary qubits and perform the correction~\cite{Li2011, Crow2016, Cruikshank2017high, Cruikshank2017role, Premakumar20}. 
For each of the 16 different values of the four-bit syndrome a unitary correction circuit would need to perform a different correction operation. This requires quite a large number of ancillas and CNOT gates as discussed in the introduction. Our room-temperature design thus represents a dramatic reduction in circuit size and duration. We also emphasize that using all-unitary processes, which is the approach we take here, provides a practical advantage; doing so avoids the need to introduce additional amplification and classical feedback circuitry.

Competing with ``active'' gate-based approaches to measurement-fee error correction, is the use of continuous autonomous  QEC~\citep{autoqecpaz1998continuous, Kapit16, Kapit17, autoqeclihm2018implementation, autoqeclescanne2020exponential}. In that family of protocols one needs to design an exotic dissipator, usually through reservoir engineering, which provides an irreversible evolution from the error-space back the code space.


\subsection*{Hardware Prospects}

We will introduce a less abstract model of our triply-resonant cavity design, in order to better describe the materials science and fabrication challenges it faces. This model also lets us give physical values for the unitless durations we have found above for our control pulses. We will start by describing the physical realization for the $\hat{a}\hat{b}^{\dagger 2}$ and $p(t)\hat{b}^\dagger\hat{c}$ terms in the Hamiltonian. Naturally, these terms requires the presence of eigenmodes $\hat{a}$, $\hat{b}$, and $\hat{c}$. The corresponding field operators would be (e.g., for the $\hat{a}$ mode)
\begin{align}
    \hat{\mathbf{B}}\_a(\mathbf{r}) & = \sqrt\frac{\hbar\omega\_a}{2}\hat{a}\mathbf{b}\_a(\mathbf{r}) + \mbox{H.c.} \\
    \hat{\mathbf{D}}\_a(\mathbf{r}) & = \sqrt\frac{\hbar\omega\_a}{2}\hat{a}\mathbf{d}\_a(\mathbf{r}) + \mbox{H.c.}
\end{align}
where we used the magnetic field and the electric displacement in order to keep the quantization consistent in the nonlinear regime~\citep{sipe2004effective,bhat2006hamiltonian,quesada2017you}. The $\mathbf{b}(\mathbf{r})$ and $\mathbf{d}(\mathbf{r})$ eigenmodes can be computed from classical electromagnetism and are normalized to
$\int\mu_0^{-1}|\mathbf{b}|^2\mathrm{d}\mathbf{r}=1$
and
$\int\varepsilon\_0^{-1}n^{-2}|\mathbf{d}|^2\mathrm{d}\mathbf{r}=1$.
The overall Hamiltonian of the system will be
\begin{equation}
\hat{H} = \int\mathrm{d}\mathbf{r}\left(
\frac{\hat{\mathbf{B}}^2}{2\mu_o}
+\frac{\hat{\mathbf{D}}^2}{2\varepsilon\_0 n^2}
-\frac{\chi^{(2)}\hat{\mathbf{D}}^3}{3\varepsilon\_0^2 n^6}
\right),
\end{equation}
where $n$ is the index of refraction (consult \citep{quesada2017you} for its complete treatment as a tensor with dispersion). The field operators are the sum of field operators for the modes $\hat{a}$, $\hat{b}$, and $\hat{c}$, as well as the field from the classical laser pulse $p(t)$. The first two terms from the Hamiltonian simply give us the harmonic oscillator terms, which we eliminate by moving to the corresponding rotating reference frames. The last term provides the nonlinear interactions in which we are interested. For simplicity, we first consider the undriven case, i.e., $p(t)=0$. The driven case is discussed in Supplementary Note 3. Expanding the nonlinear term and eliminating the non-resonant terms leaves us with
\begin{align}
\hat{H}_{\ms{nl}} & =
- \frac{\chi^{(2)}}{\sqrt{\varepsilon\_0} n^3 \sqrt{V_\text{shg}}}
~ \sqrt\frac{\hbar^3\omega\_a\omega\_b^2}{8}
~ \hat{a}\hat{b}^{\dagger 2} + \mbox{H.c.} \\
\frac{1}{\sqrt{V_{\text{shg}}}} & = \frac{
  \int_{\ms{nl}} d^{i}\_a d^{j*}\_b d^{k*}\_b \mathrm{d}\mathbf{r}}{
  \sqrt{(\int |\mathbf{d}\_b|^2 \mathrm{d}\mathbf{r})^2 \int |\mathbf{d}\_a|^2 \mathrm{d}\mathbf{r}}},
\end{align}
where $\int_{\ms{nl}}$ denotes integration only over the nonlinear medium and $i$, $j$, and $k$ denote the appropriate field components to integrate, depending on the nonlinear material being employed. Thus, $V_{\text{shg}}$ is the mode volume considered in second-harmonic generation experiments. For simplicity we are not acknowledging frequency and space dependencies in the refractive index $n$ and we are not specifying the components of the $\chi^{(2)}$ tensor being employed. This does not change the result we are pursuing.

The coupling rate in this nonlinear Hamiltonian imposes the units of time for the control pulses described in the previous section. This characteristic time needs to be compared to the cavity lifetimes, typically expressed through the $Q$ factor as $\tau = \frac{2Q}{\omega\_a}$. This lets us introduce the following figure of merit for the characteristic number of operations before the environment destroys our quantum state
\begin{equation}
    N = 
    \sqrt \frac{\hbar}{8 \varepsilon\_0} \frac{\sqrt{\omega\_a}}{n^3} \frac{ Q \chi^{(2)} }{\sqrt{V_\text{shg}}}
\end{equation}

Considering some recent second harmonic generation (SHG) on-chip experiments (a $Q\sim10^7$ in ~\citep{lin2018highly,Zhang17} and a $V_\text{shg}\sim800\mathrm{\mu m^3}\sim2000\frac{\lambda^3}{n^3}$ with a 70$\mathrm{\mu m}$-radius micro-ring in~\citep{lu2019periodically,lu2020towards}, at $\lambda\_a\approx750\mathrm{nm}$) in a typical nonlinear optics material like lithium niobate ($\chi^{(2)}\sim31\frac{\mathrm{pm}}{\mathrm{V}}$), we obtain values $N\sim0.03$, which is still too low for practical use. With $Q$ factors and mode overlaps in SHG experiments following a Moore's law (the recent progress is explored in Supplementary Note 4) and new designs lowering mode volumes by orders of magnitude~\citep{hu2016design,hu2018experimental,choi2017self,jiang2018nonlinear}, $N$ --- the number of elementary quantum operations within the cavity lifetime --- could very well grow by orders of magnitude and reach tens to hundreds over the next decade.
$N$ is related to a typical figure of merit in SHG experiments --- the conversion efficiency~\citep{guo2016second}  $\eta=\frac{P_\text{out}}{P^2_\text{in}}\propto\frac{Q^3(\chi^{(2)})^2}{V_\text{shg}}$, which has seen impressive improvements in the last decade (see Supplementary Note 4). With a $Q\sim2\times10^8$, which is achievable in principle~\citep{chris_inprep}, a mode volume of $V\sim10^{-3}\frac{\lambda^3}{n^3}$, (see~\citep{hu2018experimental,lin2016cavity} for progress in mode confinement), and $\chi^{(2)}\sim100\frac{\mathrm{pm}}{\mathrm{V}}$, which is between the values for lithium niobate and gallium arsenide, we achieve $N\sim2000$ which is enough for error correction. Moreover, new fabrication techniques for thin-film materials enable much stronger effective nonlinearities than what has otherwise been achieved on-chip. While such techniques have not been explored extensively in the optical regime, these results are an encouraging indication that similar progress may well be possible for nonlinear optical materials.

\begin{figure}
    \centering
    \includegraphics[width=3.375in]{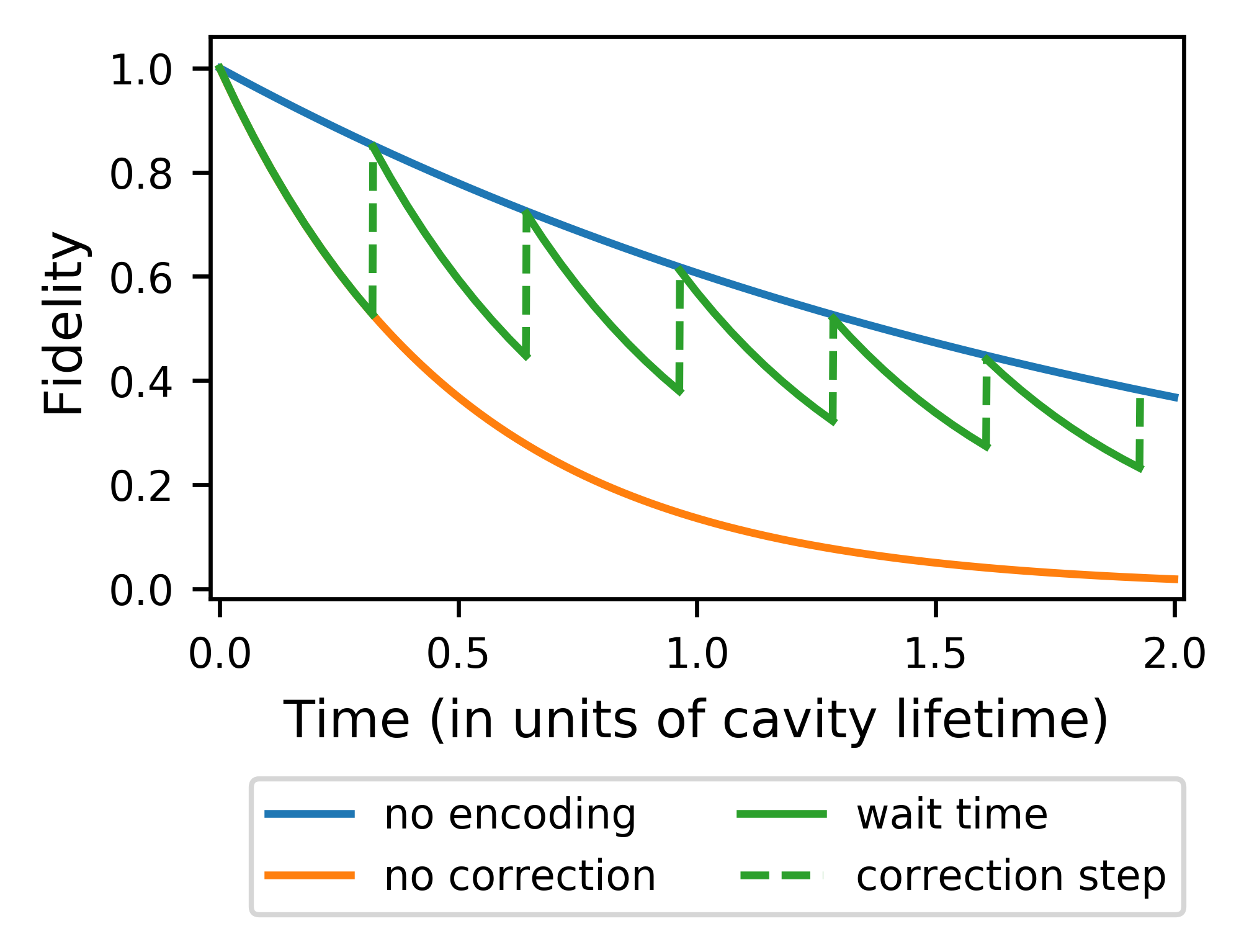}
    \caption{\emph{Logical qubit lifetime at the "break even" regime where it begins to outperform unprotected qubits.} In blue we see the decay of a single photon, i.e., an unprotected single-rail qubit. In orange we see the decay of our two-mode code if we do not perform any correction operations -- it decays faster as it contains a higher number of photons. The green line represents the decay of the encoded qubit in the presence of periodic correction operations. The infidelity of the correction operations due to photon loss that can happen during the operation is taken into account. The figure represents a lower bound for the performance of our protocol, with beneficial higher order effects being neglected in order to simplify the simulation. The "break even" point is achieved at $V_\text{shg}\sim10^{-3}\frac{\lambda^3}{n^3}$, $Q\sim2\times10^8$, and  $\chi^{(2)}\sim100\frac{\text{pm}}{\text{V}}$ for $\lambda_a\sim750\mathrm{nm}$. Waveguide losses are neglected, as they would be insignificant compared to the rest of the operations.}
    \label{fig:error_corrected_lifetime}
\end{figure}

To explore how such future hardware may perform, we compare the lifetime of an encoded (protected) photonic qubit to an unprotected single-rail qubit living in the same hardware. The time scale will be set by the $Q$ factor of the cavities under consideration, however, in order to present physical values for the parameters we will set $Q\sim2\times10^8$ at ${\lambda\_a\sim750\mathrm{nm}}$, which is well within the thermorefractive theoretical limit~\citep{chris_inprep}. In Fig.\ \ref{fig:error_corrected_lifetime} we compare the performance of our error correcting protocol to that of an unprotected single-rail qubit, and see that the error correcting threshold is $N\sim2000$, a very demanding value which we are nonetheless optimistic about given the experimental results cited earlier. Typically for non-asymptotic codes, to achieve fault tolerance this lower level code will have to be concatenated with an asymptotically growing stabilizer code, akin to the surface code or quantum LDPC codes and one of the many techniques for achieving non-Clifford gates (e.g. through magic states) will have to be employed. The versatility of our control protocol provides for a system agnostic to these higher-level architectural decisions.

Lastly, we need to consider the implementation of the time dependent control pulses. In the electrical regime, the control pulse can be modulated by standard microwave electronics in CMOS-compatible hardware~\citep{wang2018integrated}. In the optical regime the control pulse would have to be modulated by wave shaping through expressing the pulse in terms of its Fourier decomposition~\citep{weiner2011ultrafast}. Intermediate regimes are also possible, in which we can modulate a THz electric field, by placing optically-actuated Auston switches next to our triply-resonant cavities~\citep{chen2019integrated}. Active control will be necessary for loading and unloading photons from these long-lived cavities, e.g., by following methods proposed in ~\citep{heuck2019photon}.

It is important to note that one can balance the three considerations discussed in this section: the duration, power, and bandwidth of the control pulse. When the values of all these quantities can be expressed in characteristic units close to unity, the optimization problem is well conditioned and easier to solve. Such are the control pulses we have shown (e.g., their amplitudes, bandwidths, and durations are $\lesssim 10$). However, if our hardware requires short pulses (e.g., due to low $Q$ factor), but permits high power, we can nudge the solution in this direction by reparameterizing the optimization problem~\citep{methods_inprep}.

\section*{\label{sec:conclusion}Discussion}

It is accepted in the quantum computing community that any prospective purely-photonic architecture for quantum information processing would face significant challenges due to the weak photon-photon interactions available even in the best materials and resonators. Nonetheless, the present work, building upon more than a decade of theory developments on cavity-enhanced optical nonlinear interactions, shows that the monumental hardware requirements have already been nearly achieved in disparate experiments. It is an outstanding challenge to incorporate, in a single device, a record-high $Q$-factor cavity, together with extremely confined mode volumes, and fabrication-enhanced $\chi^{(2)}$ materials. However, progress over the last decade --- for example the $10^8$-fold improvement in the efficiencies of second-harmonic generation --- inspires confidence that this herculean task can very well be achieved within the next decade.

Moreover, our work, for the first time, shows that a single elementary photonic device can be reprogrammed on the fly to perform a set of diverse unitary operations, drastically lowering circuit complexity and depth. We have shown its applicability for typical single- and dual-rail encoded qubits, as well as its versatility in processing multi-photon Fock states. We showcased the flexibility of our control paradigm by devising an explicit error-correcting circuit for a bosonic code and the application of multi-qubit logic gates on top of that code. This is the first proposal for photonic logical qubits that includes compact encoding and correcting circuitry. Furthermore, the circuit we have designed does not require any measurement operations or feed-forward classical control, offering significant simplifications compared to a typical small stabilizer code, and opening the door for extremely fast, compact, room-temperature quantum repeaters.


\begin{acknowledgments}
We thank Christopher Panuski and Ryan Hamerly for many helpful conversations. Harvard Research Computing enabled much of the computational work. The SciPy, Tensorflow, and Julia open source communities provided invaluable research software. SK and MH are grateful for the funding provided by the MITRE Quantum Moonshot Programme. KJ, DE, and MH acknowledge support from a CCDC Army Research Laboratory ECI grant. 
\end{acknowledgments}

\section*{Competing Interests}

The authors declare no competing interests.

\section*{Authors' Contributions}

The design of the control protocol was performed jointly by the authors. SK wrote the optimization and analysis software and performed the simulations. The manuscript was written by SK with contributions from the other authors. 

\section*{Data Availability}

The digitized control-pulse examples in this manuscript can be readily reproduced in most optimization toolkits (e.g. Qutip and Tensorflow under Python or SciML and Flux under Julia). Upon request the authors can provide these waveforms and example scripts under each of the aforementioned frameworks that produce equivalent waveforms.

\bibliographystyle{naturemag}
\bibliography{processed_references}

\begin{thebibliography}{100}
\expandafter\ifx\csname url\endcsname\relax
  \def\url#1{\texttt{#1}}\fi
\expandafter\ifx\csname urlprefix\endcsname\relax\def\urlprefix{URL }\fi
\providecommand{\bibinfo}[2]{#2}
\providecommand{\eprint}[2][]{\url{#2}}

\bibitem{Blumoff16}
\bibinfo{author}{Blumoff, J.~Z.} \emph{et~al.}
\newblock \bibinfo{title}{Implementing and characterizing precise multiqubit
  measurements}.
\newblock \emph{\bibinfo{journal}{Phys. Rev. X}} \textbf{\bibinfo{volume}{6}},
  \bibinfo{pages}{031041} (\bibinfo{year}{2016}).
\newblock \urlprefix\url{https://link.aps.org/doi/10.1103/PhysRevX.6.031041}.

\bibitem{OMalley16}
\bibinfo{author}{O'Malley, P. J.~J.} \emph{et~al.}
\newblock \bibinfo{title}{Scalable quantum simulation of molecular energies}.
\newblock \emph{\bibinfo{journal}{Phys. Rev. X}} \textbf{\bibinfo{volume}{6}},
  \bibinfo{pages}{031007} (\bibinfo{year}{2016}).
\newblock \urlprefix\url{https://link.aps.org/doi/10.1103/PhysRevX.6.031007}.

\bibitem{Brown16}
\bibinfo{author}{Brown, K.}, \bibinfo{author}{Kim, J.} \&
  \bibinfo{author}{Monroe, C.}
\newblock \bibinfo{title}{Co-designing a scalable quantum computer with trapped
  atomic ions}.
\newblock \emph{\bibinfo{journal}{npj Quantum Inf.}}
  \textbf{\bibinfo{volume}{2}}, \bibinfo{pages}{16034} (\bibinfo{year}{2016}).

\bibitem{fu2009observation}
\bibinfo{author}{Fu, K.-M.~C.} \emph{et~al.}
\newblock \bibinfo{title}{Observation of the dynamic jahn-teller effect in the
  excited states of nitrogen-vacancy centers in diamond}.
\newblock \emph{\bibinfo{journal}{Physical Review Letters}}
  \textbf{\bibinfo{volume}{103}}, \bibinfo{pages}{256404}
  (\bibinfo{year}{2009}).

\bibitem{jahnke2015electron}
\bibinfo{author}{Jahnke, K.~D.} \emph{et~al.}
\newblock \bibinfo{title}{Electron--phonon processes of the silicon-vacancy
  centre in diamond}.
\newblock \emph{\bibinfo{journal}{New Journal of Physics}}
  \textbf{\bibinfo{volume}{17}}, \bibinfo{pages}{043011}
  (\bibinfo{year}{2015}).

\bibitem{plakhotnik2015electron}
\bibinfo{author}{Plakhotnik, T.}, \bibinfo{author}{Doherty, M.~W.} \&
  \bibinfo{author}{Manson, N.~B.}
\newblock \bibinfo{title}{Electron-phonon processes of the nitrogen-vacancy
  center in diamond}.
\newblock \emph{\bibinfo{journal}{Physical Review B}}
  \textbf{\bibinfo{volume}{92}}, \bibinfo{pages}{081203(R)}
  (\bibinfo{year}{2015}).

\bibitem{jonsson2019evaluating}
\bibinfo{author}{J{\"o}nsson, M.} \& \bibinfo{author}{Bj{\"o}rk, G.}
\newblock \bibinfo{title}{Evaluating the performance of photon-number-resolving
  detectors}.
\newblock \emph{\bibinfo{journal}{Physical Review A}}
  \textbf{\bibinfo{volume}{99}}, \bibinfo{pages}{043822}
  (\bibinfo{year}{2019}).

\bibitem{young2018general}
\bibinfo{author}{Young, S.~M.}, \bibinfo{author}{Sarovar, M.} \&
  \bibinfo{author}{L{\'e}onard, F.}
\newblock \bibinfo{title}{General modeling framework for quantum
  photodetectors}.
\newblock \emph{\bibinfo{journal}{Physical Review A}}
  \textbf{\bibinfo{volume}{98}}, \bibinfo{pages}{063835}
  (\bibinfo{year}{2018}).

\bibitem{shi2010room}
\bibinfo{author}{Shi, F.} \emph{et~al.}
\newblock \bibinfo{title}{Room-temperature implementation of the deutsch-jozsa
  algorithm with a single electronic spin in diamond}.
\newblock \emph{\bibinfo{journal}{Physical review letters}}
  \textbf{\bibinfo{volume}{105}}, \bibinfo{pages}{040504}
  (\bibinfo{year}{2010}).

\bibitem{venkataraman2013phase}
\bibinfo{author}{Venkataraman, V.}, \bibinfo{author}{Saha, K.} \&
  \bibinfo{author}{Gaeta, A.~L.}
\newblock \bibinfo{title}{Phase modulation at the few-photon level for
  weak-nonlinearity-based quantum computing}.
\newblock \emph{\bibinfo{journal}{Nature Photonics}}
  \textbf{\bibinfo{volume}{7}}, \bibinfo{pages}{138} (\bibinfo{year}{2013}).

\bibitem{nafradi2016room}
\bibinfo{author}{N{\'a}fr{\'a}di, B.}, \bibinfo{author}{Choucair, M.},
  \bibinfo{author}{Dinse, K.-P.} \& \bibinfo{author}{Forr{\'o}, L.}
\newblock \bibinfo{title}{Room temperature manipulation of long lifetime spins
  in metallic-like carbon nanospheres}.
\newblock \emph{\bibinfo{journal}{Nature communications}}
  \textbf{\bibinfo{volume}{7}}, \bibinfo{pages}{1--8} (\bibinfo{year}{2016}).

\bibitem{ivady2019stabilization}
\bibinfo{author}{Iv{\'a}dy, V.} \emph{et~al.}
\newblock \bibinfo{title}{Stabilization of point-defect spin qubits by quantum
  wells}.
\newblock \emph{\bibinfo{journal}{Nature communications}}
  \textbf{\bibinfo{volume}{10}}, \bibinfo{pages}{1--8} (\bibinfo{year}{2019}).

\bibitem{shapiro2006single}
\bibinfo{author}{Shapiro, J.~H.}
\newblock \bibinfo{title}{Single-photon {K}err nonlinearities do not help
  quantum computation}.
\newblock \emph{\bibinfo{journal}{Physical Review A}}
  \textbf{\bibinfo{volume}{73}}, \bibinfo{pages}{062305}
  (\bibinfo{year}{2006}).

\bibitem{leung2009spectral}
\bibinfo{author}{Leung, P.~M.}, \bibinfo{author}{Munro, W.~J.},
  \bibinfo{author}{Nemoto, K.} \& \bibinfo{author}{Ralph, T.~C.}
\newblock \bibinfo{title}{Spectral effects of strong $\chi$ (2) nonlinearity
  for quantum processing}.
\newblock \emph{\bibinfo{journal}{Physical Review A}}
  \textbf{\bibinfo{volume}{79}}, \bibinfo{pages}{042307}
  (\bibinfo{year}{2009}).

\bibitem{gea2010impossibility}
\bibinfo{author}{Gea-Banacloche, J.}
\newblock \bibinfo{title}{Impossibility of large phase shifts via the giant
  kerr effect with single-photon wave packets}.
\newblock \emph{\bibinfo{journal}{Physical Review A}}
  \textbf{\bibinfo{volume}{81}}, \bibinfo{pages}{043823}
  (\bibinfo{year}{2010}).

\bibitem{he2011transverse}
\bibinfo{author}{He, B.}, \bibinfo{author}{MacRae, A.}, \bibinfo{author}{Han,
  Y.}, \bibinfo{author}{Lvovsky, A.~I.} \& \bibinfo{author}{Simon, C.}
\newblock \bibinfo{title}{Transverse multimode effects on the performance of
  photon-photon gates}.
\newblock \emph{\bibinfo{journal}{Physical Review A}}
  \textbf{\bibinfo{volume}{83}}, \bibinfo{pages}{022312}
  (\bibinfo{year}{2011}).

\bibitem{xu2013analytic}
\bibinfo{author}{Xu, S.}, \bibinfo{author}{Rephaeli, E.} \&
  \bibinfo{author}{Fan, S.}
\newblock \bibinfo{title}{Analytic properties of two-photon scattering matrix
  in integrated quantum systems determined by the cluster decomposition
  principle}.
\newblock \emph{\bibinfo{journal}{Physical review letters}}
  \textbf{\bibinfo{volume}{111}}, \bibinfo{pages}{223602}
  (\bibinfo{year}{2013}).

\bibitem{dove2014phase}
\bibinfo{author}{Dove, J.}, \bibinfo{author}{Chudzicki, C.} \&
  \bibinfo{author}{Shapiro, J.~H.}
\newblock \bibinfo{title}{Phase-noise limitations on single-photon cross-phase
  modulation with differing group velocities}.
\newblock \emph{\bibinfo{journal}{Physical Review A}}
  \textbf{\bibinfo{volume}{90}}, \bibinfo{pages}{062314}
  (\bibinfo{year}{2014}).

\bibitem{hu2016design}
\bibinfo{author}{Hu, S.} \& \bibinfo{author}{Weiss, S.~M.}
\newblock \bibinfo{title}{Design of photonic crystal cavities for extreme light
  concentration}.
\newblock \emph{\bibinfo{journal}{ACS photonics}} \textbf{\bibinfo{volume}{3}},
  \bibinfo{pages}{1647--1653} (\bibinfo{year}{2016}).

\bibitem{hu2018experimental}
\bibinfo{author}{Hu, S.} \emph{et~al.}
\newblock \bibinfo{title}{Experimental realization of deep-subwavelength
  confinement in dielectric optical resonators}.
\newblock \emph{\bibinfo{journal}{Science advances}}
  \textbf{\bibinfo{volume}{4}}, \bibinfo{pages}{eaat2355}
  (\bibinfo{year}{2018}).

\bibitem{choi2017self}
\bibinfo{author}{Choi, H.}, \bibinfo{author}{Heuck, M.} \&
  \bibinfo{author}{Englund, D.}
\newblock \bibinfo{title}{Self-similar nanocavity design with ultrasmall mode
  volume for single-photon nonlinearities}.
\newblock \emph{\bibinfo{journal}{Physical review letters}}
  \textbf{\bibinfo{volume}{118}}, \bibinfo{pages}{223605}
  (\bibinfo{year}{2017}).

\bibitem{heuck2019controlled}
\bibinfo{author}{Heuck, M.}, \bibinfo{author}{Jacobs, K.} \&
  \bibinfo{author}{Englund, D.~R.}
\newblock \bibinfo{title}{Controlled-phase gate using dynamically coupled
  cavities and optical nonlinearities}.
\newblock \emph{\bibinfo{journal}{arXiv preprint arXiv:1909.05751}}
  (\bibinfo{year}{2019}).

\bibitem{heuck2019photon}
\bibinfo{author}{Heuck, M.}, \bibinfo{author}{Jacobs, K.} \&
  \bibinfo{author}{Englund, D.~R.}
\newblock \bibinfo{title}{Photon-photon interactions in dynamically coupled
  cavities}.
\newblock \emph{\bibinfo{journal}{arXiv preprint arXiv:1905.02134}}
  (\bibinfo{year}{2019}).

\bibitem{li2019photon}
\bibinfo{author}{Li, M.} \emph{et~al.}
\newblock \bibinfo{title}{Photon-photon quantum phase gate in a photonic
  molecule with $\chi^{2}$ nonlinearity}.
\newblock \emph{\bibinfo{journal}{arXiv preprint arXiv:1909.10839}}
  (\bibinfo{year}{2019}).

\bibitem{Kok07}
\bibinfo{author}{Kok, P.} \emph{et~al.}
\newblock \bibinfo{title}{Linear optical quantum computing with photonic
  qubits}.
\newblock \emph{\bibinfo{journal}{Rev. Mod. Phys.}}
  \textbf{\bibinfo{volume}{79}}, \bibinfo{pages}{135--174}
  (\bibinfo{year}{2007}).
\newblock \urlprefix\url{https://link.aps.org/doi/10.1103/RevModPhys.79.135}.

\bibitem{Laflamme96}
\bibinfo{author}{Laflamme, R.}, \bibinfo{author}{Miquel, C.},
  \bibinfo{author}{Paz, J.~P.} \& \bibinfo{author}{Zurek, W.~H.}
\newblock \bibinfo{title}{Perfect quantum error correcting code}.
\newblock \emph{\bibinfo{journal}{Phys. Rev. Lett.}}
  \textbf{\bibinfo{volume}{77}}, \bibinfo{pages}{198--201}
  (\bibinfo{year}{1996}).
\newblock \urlprefix\url{https://link.aps.org/doi/10.1103/PhysRevLett.77.198}.

\bibitem{divincenzo1996fault}
\bibinfo{author}{DiVincenzo, D.~P.} \& \bibinfo{author}{Shor, P.~W.}
\newblock \bibinfo{title}{Fault-tolerant error correction with efficient
  quantum codes}.
\newblock \emph{\bibinfo{journal}{Physical review letters}}
  \textbf{\bibinfo{volume}{77}}, \bibinfo{pages}{3260} (\bibinfo{year}{1996}).

\bibitem{Crow2016}
\bibinfo{author}{Crow, D.}, \bibinfo{author}{Joynt, R.} \&
  \bibinfo{author}{Saffman, M.}
\newblock \bibinfo{title}{Improved error thresholds for measurement-free error
  correction}.
\newblock \emph{\bibinfo{journal}{Phys. Rev. Lett.}}
  \textbf{\bibinfo{volume}{117}}, \bibinfo{pages}{130503}
  (\bibinfo{year}{2016}).
\newblock
  \urlprefix\url{https://link.aps.org/doi/10.1103/PhysRevLett.117.130503}.

\bibitem{Premakumar20}
\bibinfo{author}{Premakumar, V.~N.}, \bibinfo{author}{Saffman, M.} \&
  \bibinfo{author}{Joynt, R.}
\newblock \bibinfo{title}{Measurement-free error correction with coherent
  ancillas}.
\newblock \emph{\bibinfo{journal}{Preprint: ArXiv:2007.09804}}
  (\bibinfo{year}{2020}).

\bibitem{niu2018qudit}
\bibinfo{author}{Niu, M.~Y.}, \bibinfo{author}{Chuang, I.~L.} \&
  \bibinfo{author}{Shapiro, J.~H.}
\newblock \bibinfo{title}{Qudit-basis universal quantum computation using
  $\chi$ (2) interactions}.
\newblock \emph{\bibinfo{journal}{Physical review letters}}
  \textbf{\bibinfo{volume}{120}}, \bibinfo{pages}{160502}
  (\bibinfo{year}{2018}).

\bibitem{niu2018hardware}
\bibinfo{author}{Niu, M.~Y.}, \bibinfo{author}{Chuang, I.~L.} \&
  \bibinfo{author}{Shapiro, J.~H.}
\newblock \bibinfo{title}{Hardware-efficient bosonic quantum error-correcting
  codes based on symmetry operators}.
\newblock \emph{\bibinfo{journal}{Physical Review A}}
  \textbf{\bibinfo{volume}{97}}, \bibinfo{pages}{032323}
  (\bibinfo{year}{2018}).

\bibitem{Michael16}
\bibinfo{author}{Michael, M.~H.} \emph{et~al.}
\newblock \bibinfo{title}{New class of quantum error-correcting codes for a
  bosonic mode}.
\newblock \emph{\bibinfo{journal}{Phys. Rev. X}} \textbf{\bibinfo{volume}{6}},
  \bibinfo{pages}{031006} (\bibinfo{year}{2016}).
\newblock \urlprefix\url{https://link.aps.org/doi/10.1103/PhysRevX.6.031006}.

\bibitem{Albert18}
\bibinfo{author}{Albert, V.~V.} \emph{et~al.}
\newblock \bibinfo{title}{Performance and structure of single-mode bosonic
  codes}.
\newblock \emph{\bibinfo{journal}{Phys. Rev. A}} \textbf{\bibinfo{volume}{97}},
  \bibinfo{pages}{032346} (\bibinfo{year}{2018}).
\newblock \urlprefix\url{https://link.aps.org/doi/10.1103/PhysRevA.97.032346}.

\bibitem{bergmann2016quantum}
\bibinfo{author}{Bergmann, M.} \& \bibinfo{author}{van Loock, P.}
\newblock \bibinfo{title}{Quantum error correction against photon loss using
  {N}{O}{O}{N} states}.
\newblock \emph{\bibinfo{journal}{Physical Review A}}
  \textbf{\bibinfo{volume}{94}}, \bibinfo{pages}{012311}
  (\bibinfo{year}{2016}).

\bibitem{steinbrecher2019quantum}
\bibinfo{author}{Steinbrecher, G.~R.}, \bibinfo{author}{Olson, J.~P.},
  \bibinfo{author}{Englund, D.} \& \bibinfo{author}{Carolan, J.}
\newblock \bibinfo{title}{Quantum optical neural networks}.
\newblock \emph{\bibinfo{journal}{npj Quantum Information}}
  \textbf{\bibinfo{volume}{5}}, \bibinfo{pages}{1--9} (\bibinfo{year}{2019}).

\bibitem{weedbrook2012gaussian}
\bibinfo{author}{Weedbrook, C.} \emph{et~al.}
\newblock \bibinfo{title}{Gaussian quantum information}.
\newblock \emph{\bibinfo{journal}{Reviews of Modern Physics}}
  \textbf{\bibinfo{volume}{84}}, \bibinfo{pages}{621} (\bibinfo{year}{2012}).

\bibitem{bartlett2002efficient}
\bibinfo{author}{Bartlett, S.~D.}, \bibinfo{author}{Sanders, B.~C.},
  \bibinfo{author}{Braunstein, S.~L.} \& \bibinfo{author}{Nemoto, K.}
\newblock \bibinfo{title}{Efficient classical simulation of continuous variable
  quantum information processes}.
\newblock \emph{\bibinfo{journal}{Physical Review Letters}}
  \textbf{\bibinfo{volume}{88}}, \bibinfo{pages}{097904}
  (\bibinfo{year}{2002}).

\bibitem{lloyd1999quantum}
\bibinfo{author}{Lloyd, S.} \& \bibinfo{author}{Braunstein, S.~L.}
\newblock \bibinfo{title}{Quantum computation over continuous variables}.
\newblock In \emph{\bibinfo{booktitle}{Quantum Information with Continuous
  Variables}}, \bibinfo{pages}{9--17} (\bibinfo{publisher}{Springer},
  \bibinfo{year}{1999}).
\newblock \eprint{quant-ph/9810082}.

\bibitem{jacobs2007engineering}
\bibinfo{author}{Jacobs, K.}
\newblock \bibinfo{title}{Engineering quantum states of a nanoresonator via a
  simple auxiliary system}.
\newblock \emph{\bibinfo{journal}{Physical review letters}}
  \textbf{\bibinfo{volume}{99}}, \bibinfo{pages}{117203}
  (\bibinfo{year}{2007}).

\bibitem{krastanov2015universal}
\bibinfo{author}{Krastanov, S.} \emph{et~al.}
\newblock \bibinfo{title}{Universal control of an oscillator with dispersive
  coupling to a qubit}.
\newblock \emph{\bibinfo{journal}{Physical Review A}}
  \textbf{\bibinfo{volume}{92}}, \bibinfo{pages}{040303(R)}
  (\bibinfo{year}{2015}).

\bibitem{methods_inprep}
\bibinfo{author}{Krastanov, S.}
\newblock \bibinfo{title}{Symbolic and numerical tools for quantum optimal
  control} (\bibinfo{year}{in prep.}).

\bibitem{khaneja2005optimal}
\bibinfo{author}{Khaneja, N.}, \bibinfo{author}{Reiss, T.},
  \bibinfo{author}{Kehlet, C.}, \bibinfo{author}{Schulte-Herbr{\"u}ggen, T.} \&
  \bibinfo{author}{Glaser, S.~J.}
\newblock \bibinfo{title}{Optimal control of coupled spin dynamics: design of
  nmr pulse sequences by gradient ascent algorithms}.
\newblock \emph{\bibinfo{journal}{Journal of magnetic resonance}}
  \textbf{\bibinfo{volume}{172}}, \bibinfo{pages}{296--305}
  (\bibinfo{year}{2005}).

\bibitem{heeres2017implementing}
\bibinfo{author}{Heeres, R.~W.} \emph{et~al.}
\newblock \bibinfo{title}{Implementing a universal gate set on a logical qubit
  encoded in an oscillator}.
\newblock \emph{\bibinfo{journal}{Nature communications}}
  \textbf{\bibinfo{volume}{8}}, \bibinfo{pages}{94} (\bibinfo{year}{2017}).

\bibitem{shi2002both}
\bibinfo{author}{Shi, Y.}
\newblock \bibinfo{title}{Both toffoli and controlled-not need little help to
  do universal quantum computation}.
\newblock \emph{\bibinfo{journal}{arXiv preprint quant-ph/0205115}}
  (\bibinfo{year}{2002}).

\bibitem{aharonov2003simple}
\bibinfo{author}{Aharonov, D.}
\newblock \bibinfo{title}{A simple proof that toffoli and hadamard are quantum
  universal}.
\newblock \emph{\bibinfo{journal}{arXiv preprint quant-ph/0301040}}
  (\bibinfo{year}{2003}).

\bibitem{barenco1995elementary}
\bibinfo{author}{Barenco, A.} \emph{et~al.}
\newblock \bibinfo{title}{Elementary gates for quantum computation}.
\newblock \emph{\bibinfo{journal}{Physical review A}}
  \textbf{\bibinfo{volume}{52}}, \bibinfo{pages}{3457} (\bibinfo{year}{1995}).

\bibitem{shende2008cnot}
\bibinfo{author}{Shende, V.~V.} \& \bibinfo{author}{Markov, I.~L.}
\newblock \bibinfo{title}{On the cnot-cost of toffoli gates}.
\newblock \emph{\bibinfo{journal}{arXiv preprint arXiv:0803.2316}}
  (\bibinfo{year}{2008}).

\bibitem{lloyd2000coherent}
\bibinfo{author}{Lloyd, S.}
\newblock \bibinfo{title}{Coherent quantum feedback}.
\newblock \emph{\bibinfo{journal}{Physical Review A}}
  \textbf{\bibinfo{volume}{62}}, \bibinfo{pages}{022108}
  (\bibinfo{year}{2000}).

\bibitem{Jacobs2014coherent}
\bibinfo{author}{Jacobs, K.}, \bibinfo{author}{Wang, X.} \&
  \bibinfo{author}{Wiseman, H.~M.}
\newblock \bibinfo{title}{Coherent feedback that beats all measurement-based
  feedback protocols}.
\newblock \emph{\bibinfo{journal}{New Journal of Physics}}
  \textbf{\bibinfo{volume}{16}}, \bibinfo{pages}{073036}
  (\bibinfo{year}{2014}).
\newblock
  \urlprefix\url{https://doi.org/10.1088%2F1367-2630%2F16%2F7%2F073036}.

\bibitem{carolan2015universal}
\bibinfo{author}{Carolan, J.} \emph{et~al.}
\newblock \bibinfo{title}{Universal linear optics}.
\newblock \emph{\bibinfo{journal}{Science}} \textbf{\bibinfo{volume}{349}},
  \bibinfo{pages}{711--716} (\bibinfo{year}{2015}).

\bibitem{Li2011}
\bibinfo{author}{Li, C.-K.}, \bibinfo{author}{Nakahara, M.},
  \bibinfo{author}{Poon, Y.}, \bibinfo{author}{Sze, N.-S.} \&
  \bibinfo{author}{Tomita, H.}
\newblock \bibinfo{title}{Recovery in quantum error correction for general
  noise without measurement}.
\newblock \emph{\bibinfo{journal}{Quantum Inf. Comput.}}
  \textbf{\bibinfo{volume}{12}}, \bibinfo{pages}{0149} (\bibinfo{year}{2011}).

\bibitem{Cruikshank2017high}
\bibinfo{author}{Cruikshank, B.} \& \bibinfo{author}{Jacobs, K.}
\newblock \bibinfo{title}{High-threshold low-overhead fault-tolerant classical
  computation and the replacement of measurements with unitary quantum gates}.
\newblock \emph{\bibinfo{journal}{Phys. Rev. Lett.}}
  \textbf{\bibinfo{volume}{119}}, \bibinfo{pages}{030503}
  (\bibinfo{year}{2017}).
\newblock
  \urlprefix\url{https://link.aps.org/doi/10.1103/PhysRevLett.119.030503}.

\bibitem{Cruikshank2017role}
\bibinfo{author}{Cruikshank, B.} \& \bibinfo{author}{Jacobs, K.}
\newblock \bibinfo{title}{The role of quantum measurements in physical
  processes and protocols}.
\newblock \emph{\bibinfo{journal}{Quantum Science and Technology}}
  \textbf{\bibinfo{volume}{2}}, \bibinfo{pages}{033001} (\bibinfo{year}{2017}).

\bibitem{autoqecpaz1998continuous}
\bibinfo{author}{Paz, J.~P.} \& \bibinfo{author}{Zurek, W.~H.}
\newblock \bibinfo{title}{Continuous error correction}.
\newblock \emph{\bibinfo{journal}{Proceedings of the Royal Society of London.
  Series A: Mathematical, Physical and Engineering Sciences}}
  \textbf{\bibinfo{volume}{454}}, \bibinfo{pages}{355--364}
  (\bibinfo{year}{1998}).

\bibitem{Kapit16}
\bibinfo{author}{Kapit, E.}
\newblock \bibinfo{title}{Hardware-efficient and fully autonomous quantum error
  correction in superconducting circuits}.
\newblock \emph{\bibinfo{journal}{Phys. Rev. Lett.}}
  \textbf{\bibinfo{volume}{116}}, \bibinfo{pages}{150501}
  (\bibinfo{year}{2016}).
\newblock
  \urlprefix\url{https://link.aps.org/doi/10.1103/PhysRevLett.116.150501}.

\bibitem{Kapit17}
\bibinfo{author}{Kapit, E.}
\newblock \bibinfo{title}{The upside of noise: engineered dissipation as a
  resource in superconducting circuits}.
\newblock \emph{\bibinfo{journal}{Quantum Science and Technology}}
  \bibinfo{pages}{033002} (\bibinfo{year}{2017}).

\bibitem{autoqeclihm2018implementation}
\bibinfo{author}{Lihm, J.-M.}, \bibinfo{author}{Noh, K.} \&
  \bibinfo{author}{Fischer, U.~R.}
\newblock \bibinfo{title}{Implementation-independent sufficient condition of
  the {K}nill-{L}aflamme type for the autonomous protection of logical qudits
  by strong engineered dissipation}.
\newblock \emph{\bibinfo{journal}{Physical Review A}}
  \textbf{\bibinfo{volume}{98}}, \bibinfo{pages}{012317}
  (\bibinfo{year}{2018}).

\bibitem{autoqeclescanne2020exponential}
\bibinfo{author}{Lescanne, R.} \emph{et~al.}
\newblock \bibinfo{title}{Exponential suppression of bit-flips in a qubit
  encoded in an oscillator}.
\newblock \emph{\bibinfo{journal}{Nature Physics}}
  \textbf{\bibinfo{volume}{16}}, \bibinfo{pages}{509--513}
  (\bibinfo{year}{2020}).

\bibitem{sipe2004effective}
\bibinfo{author}{Sipe, J.~E.}, \bibinfo{author}{Bhat, N. A.~R.},
  \bibinfo{author}{Chak, P.} \& \bibinfo{author}{Pereira, S.}
\newblock \bibinfo{title}{Effective field theory for the nonlinear optical
  properties of photonic crystals}.
\newblock \emph{\bibinfo{journal}{Physical Review E}}
  \textbf{\bibinfo{volume}{69}}, \bibinfo{pages}{016604}
  (\bibinfo{year}{2004}).

\bibitem{bhat2006hamiltonian}
\bibinfo{author}{Bhat, N. A.~R.} \& \bibinfo{author}{Sipe, J.~E.}
\newblock \bibinfo{title}{Hamiltonian treatment of the electromagnetic field in
  dispersive and absorptive structured media}.
\newblock \emph{\bibinfo{journal}{Physical Review A}}
  \textbf{\bibinfo{volume}{73}}, \bibinfo{pages}{063808}
  (\bibinfo{year}{2006}).

\bibitem{quesada2017you}
\bibinfo{author}{Quesada, N.} \& \bibinfo{author}{Sipe, J.}
\newblock \bibinfo{title}{Why you should not use the electric field to quantize
  in nonlinear optics}.
\newblock \emph{\bibinfo{journal}{Optics letters}}
  \textbf{\bibinfo{volume}{42}}, \bibinfo{pages}{3443--3446}
  (\bibinfo{year}{2017}).

\bibitem{lin2018highly}
\bibinfo{author}{Lin, J.} \emph{et~al.}
\newblock \bibinfo{title}{Highly-efficient second and third harmonic generation
  in a monocrystalline lithium niobate microresonator}.
\newblock \emph{\bibinfo{journal}{arXiv preprint arXiv:1809.04523}}
  (\bibinfo{year}{2018}).

\bibitem{Zhang17}
\bibinfo{author}{Zhang, M.}, \bibinfo{author}{Wang, C.},
  \bibinfo{author}{Cheng, R.}, \bibinfo{author}{Shams-Ansari, A.} \&
  \bibinfo{author}{Lon\v{c}ar, M.}
\newblock \bibinfo{title}{Monolithic ultra-high-q lithium niobate microring
  resonator}.
\newblock \emph{\bibinfo{journal}{Optica}} \textbf{\bibinfo{volume}{4}},
  \bibinfo{pages}{1536--1537} (\bibinfo{year}{2017}).
\newblock
  \urlprefix\url{http://www.osapublishing.org/optica/abstract.cfm?URI=optica-4-12-1536}.

\bibitem{lu2019periodically}
\bibinfo{author}{Lu, J.} \emph{et~al.}
\newblock \bibinfo{title}{Periodically poled thin-film lithium niobate
  microring resonators with a second-harmonic generation efficiency of
  250,000\%/w}.
\newblock \emph{\bibinfo{journal}{Optica}} \textbf{\bibinfo{volume}{6}},
  \bibinfo{pages}{1455--1460} (\bibinfo{year}{2019}).

\bibitem{lu2020towards}
\bibinfo{author}{Lu, J.}, \bibinfo{author}{Li, M.}, \bibinfo{author}{Zou,
  C.-L.}, \bibinfo{author}{Sayem, A.~A.} \& \bibinfo{author}{Tang, H.~X.}
\newblock \bibinfo{title}{Towards 1\% single photon nonlinearity with
  periodically-poled lithium niobate microring resonators}.
\newblock \emph{\bibinfo{journal}{arXiv preprint arXiv:2007.07411}}
  (\bibinfo{year}{2020}).

\bibitem{jiang2018nonlinear}
\bibinfo{author}{Jiang, H.} \emph{et~al.}
\newblock \bibinfo{title}{Nonlinear frequency conversion in one dimensional
  lithium niobate photonic crystal nanocavities}.
\newblock \emph{\bibinfo{journal}{Applied Physics Letters}}
  \textbf{\bibinfo{volume}{113}}, \bibinfo{pages}{021104}
  (\bibinfo{year}{2018}).

\bibitem{guo2016second}
\bibinfo{author}{Guo, X.}, \bibinfo{author}{Zou, C.-L.} \&
  \bibinfo{author}{Tang, H.~X.}
\newblock \bibinfo{title}{Second-harmonic generation in aluminum nitride
  microrings with 2500\%/w conversion efficiency}.
\newblock \emph{\bibinfo{journal}{Optica}} \textbf{\bibinfo{volume}{3}},
  \bibinfo{pages}{1126--1131} (\bibinfo{year}{2016}).

\bibitem{chris_inprep}
\bibinfo{author}{Panuski, C.}, \bibinfo{author}{Hamerly, R.} \&
  \bibinfo{author}{Englund, D.}
\newblock \bibinfo{title}{Fundamental thermal noise limits for high-q/v optical
  microcavities}.
\newblock \emph{\bibinfo{journal}{Bulletin of the American Physical Society}}
  (\bibinfo{year}{2020}).

\bibitem{lin2016cavity}
\bibinfo{author}{Lin, Z.}, \bibinfo{author}{Liang, X.},
  \bibinfo{author}{Lon{\v{c}}ar, M.}, \bibinfo{author}{Johnson, S.~G.} \&
  \bibinfo{author}{Rodriguez, A.~W.}
\newblock \bibinfo{title}{Cavity-enhanced second-harmonic generation via
  nonlinear-overlap optimization}.
\newblock \emph{\bibinfo{journal}{Optica}} \textbf{\bibinfo{volume}{3}},
  \bibinfo{pages}{233--238} (\bibinfo{year}{2016}).

\bibitem{wang2018integrated}
\bibinfo{author}{Wang, C.} \emph{et~al.}
\newblock \bibinfo{title}{Integrated lithium niobate electro-optic modulators
  operating at cmos-compatible voltages}.
\newblock \emph{\bibinfo{journal}{Nature}} \textbf{\bibinfo{volume}{562}},
  \bibinfo{pages}{101--104} (\bibinfo{year}{2018}).

\bibitem{weiner2011ultrafast}
\bibinfo{author}{Weiner, A.~M.}
\newblock \bibinfo{title}{Ultrafast optical pulse shaping: A tutorial review}.
\newblock \emph{\bibinfo{journal}{Optics Communications}}
  \textbf{\bibinfo{volume}{284}}, \bibinfo{pages}{3669--3692}
  (\bibinfo{year}{2011}).

\bibitem{chen2019integrated}
\bibinfo{author}{Chen, P.}, \bibinfo{author}{Hosseini, M.} \&
  \bibinfo{author}{Babakhani, A.}
\newblock \bibinfo{title}{An integrated germanium-based thz impulse radiator
  with an optical waveguide coupled photoconductive switch in silicon}.
\newblock \emph{\bibinfo{journal}{Micromachines}}
  \textbf{\bibinfo{volume}{10}}, \bibinfo{pages}{367} (\bibinfo{year}{2019}).

\bibitem{ducci2004continuous}
\bibinfo{author}{Ducci, S.} \emph{et~al.}
\newblock \bibinfo{title}{Continuous-wave second-harmonic generation in modal
  phase matched semiconductor waveguides}.
\newblock \emph{\bibinfo{journal}{Applied physics letters}}
  \textbf{\bibinfo{volume}{84}}, \bibinfo{pages}{2974--2976}
  (\bibinfo{year}{2004}).

\bibitem{yu2005efficient}
\bibinfo{author}{Yu, X.}, \bibinfo{author}{Scaccabarozzi, L.},
  \bibinfo{author}{Harris, J.}, \bibinfo{author}{Kuo, P.} \&
  \bibinfo{author}{Fejer, M.}
\newblock \bibinfo{title}{Efficient continuous wave second harmonic generation
  pumped at 1.55 $\mu$m in quasi-phase-matched algaas waveguides}.
\newblock \emph{\bibinfo{journal}{Optics Express}}
  \textbf{\bibinfo{volume}{13}}, \bibinfo{pages}{10742--10748}
  (\bibinfo{year}{2005}).

\bibitem{galli2010low}
\bibinfo{author}{Galli, M.} \emph{et~al.}
\newblock \bibinfo{title}{Low-power continuous-wave generation of visible
  harmonics in silicon photonic crystal nanocavities}.
\newblock \emph{\bibinfo{journal}{Optics express}}
  \textbf{\bibinfo{volume}{18}}, \bibinfo{pages}{26613--26624}
  (\bibinfo{year}{2010}).

\bibitem{xiong2011integrated}
\bibinfo{author}{Xiong, C.} \emph{et~al.}
\newblock \bibinfo{title}{Integrated gan photonic circuits on silicon (100) for
  second harmonic generation}.
\newblock \emph{\bibinfo{journal}{Optics express}}
  \textbf{\bibinfo{volume}{19}}, \bibinfo{pages}{10462--10470}
  (\bibinfo{year}{2011}).

\bibitem{wang2012high}
\bibinfo{author}{Wang, T.-J.}, \bibinfo{author}{He, J.-Y.},
  \bibinfo{author}{Lee, C.-A.} \& \bibinfo{author}{Niu, H.}
\newblock \bibinfo{title}{High-quality linbo 3 microdisk resonators by undercut
  etching and surface tension reshaping}.
\newblock \emph{\bibinfo{journal}{Optics Express}}
  \textbf{\bibinfo{volume}{20}}, \bibinfo{pages}{28119--28124}
  (\bibinfo{year}{2012}).

\bibitem{diziain2013second}
\bibinfo{author}{Diziain, S.} \emph{et~al.}
\newblock \bibinfo{title}{Second harmonic generation in free-standing lithium
  niobate photonic crystal l3 cavity}.
\newblock \emph{\bibinfo{journal}{Applied Physics Letters}}
  \textbf{\bibinfo{volume}{103}}, \bibinfo{pages}{051117}
  (\bibinfo{year}{2013}).

\bibitem{lengle2013efficient}
\bibinfo{author}{Lengl{\'e}, K.} \emph{et~al.}
\newblock \bibinfo{title}{Efficient second harmonic generation in nanophotonic
  waveguides for optical signal processing}.
\newblock \emph{\bibinfo{journal}{Applied Physics Letters}}
  \textbf{\bibinfo{volume}{102}}, \bibinfo{pages}{151114}
  (\bibinfo{year}{2013}).

\bibitem{kuo2014second}
\bibinfo{author}{Kuo, P.~S.}, \bibinfo{author}{Bravo-Abad, J.} \&
  \bibinfo{author}{Solomon, G.~S.}
\newblock \bibinfo{title}{Second-harmonic generation using-quasi-phasematching
  in a gaas whispering-gallery-mode microcavity}.
\newblock \emph{\bibinfo{journal}{Nature communications}}
  \textbf{\bibinfo{volume}{5}}, \bibinfo{pages}{3109} (\bibinfo{year}{2014}).

\bibitem{wang2014integrated}
\bibinfo{author}{Wang, C.} \emph{et~al.}
\newblock \bibinfo{title}{Integrated high quality factor lithium niobate
  microdisk resonators}.
\newblock \emph{\bibinfo{journal}{Optics express}}
  \textbf{\bibinfo{volume}{22}}, \bibinfo{pages}{30924--30933}
  (\bibinfo{year}{2014}).

\bibitem{lin2015fabrication}
\bibinfo{author}{Lin, J.} \emph{et~al.}
\newblock \bibinfo{title}{Fabrication of high-q lithium niobate microresonators
  using femtosecond laser micromachining}.
\newblock \emph{\bibinfo{journal}{Scientific reports}}
  \textbf{\bibinfo{volume}{5}}, \bibinfo{pages}{8072} (\bibinfo{year}{2015}).

\bibitem{wang2015high}
\bibinfo{author}{Wang, J.} \emph{et~al.}
\newblock \bibinfo{title}{High-q lithium niobate microdisk resonators on a chip
  for efficient electro-optic modulation}.
\newblock \emph{\bibinfo{journal}{Optics express}}
  \textbf{\bibinfo{volume}{23}}, \bibinfo{pages}{23072--23078}
  (\bibinfo{year}{2015}).

\bibitem{moore2016efficient}
\bibinfo{author}{Moore, J.} \emph{et~al.}
\newblock \bibinfo{title}{Efficient second harmonic generation in lithium
  niobate on insulator}.
\newblock In \emph{\bibinfo{booktitle}{2016 Conference on Lasers and
  Electro-Optics (CLEO)}}, \bibinfo{pages}{1--2} (\bibinfo{organization}{IEEE},
  \bibinfo{year}{2016}).

\bibitem{lin2016phase}
\bibinfo{author}{Lin, J.} \emph{et~al.}
\newblock \bibinfo{title}{Phase-matched second-harmonic generation in an
  on-chip l i nbo 3 microresonator}.
\newblock \emph{\bibinfo{journal}{Physical Review Applied}}
  \textbf{\bibinfo{volume}{6}}, \bibinfo{pages}{014002} (\bibinfo{year}{2016}).

\bibitem{luo2017chip}
\bibinfo{author}{Luo, R.} \emph{et~al.}
\newblock \bibinfo{title}{On-chip second-harmonic generation and broadband
  parametric down-conversion in a lithium niobate microresonator}.
\newblock \emph{\bibinfo{journal}{Optics express}}
  \textbf{\bibinfo{volume}{25}}, \bibinfo{pages}{24531--24539}
  (\bibinfo{year}{2017}).

\bibitem{wang2018high}
\bibinfo{author}{Wang, L.} \emph{et~al.}
\newblock \bibinfo{title}{High-q chaotic lithium niobate microdisk cavity}.
\newblock \emph{\bibinfo{journal}{Optics letters}}
  \textbf{\bibinfo{volume}{43}}, \bibinfo{pages}{2917--2920}
  (\bibinfo{year}{2018}).

\bibitem{wolf2018quasi}
\bibinfo{author}{Wolf, R.} \emph{et~al.}
\newblock \bibinfo{title}{Quasi-phase-matched nonlinear optical frequency
  conversion in on-chip whispering galleries}.
\newblock \emph{\bibinfo{journal}{Optica}} \textbf{\bibinfo{volume}{5}},
  \bibinfo{pages}{872--875} (\bibinfo{year}{2018}).

\bibitem{li2018high}
\bibinfo{author}{Li, M.}, \bibinfo{author}{Liang, H.}, \bibinfo{author}{Luo,
  R.}, \bibinfo{author}{He, Y.} \& \bibinfo{author}{Lin, Q.}
\newblock \bibinfo{title}{High-q two-dimensional lithium niobate photonic
  crystal slab nanoresonators}.
\newblock \emph{\bibinfo{journal}{arXiv preprint arXiv:1806.04755}}
  (\bibinfo{year}{2018}).

\bibitem{luo2019optical}
\bibinfo{author}{Luo, R.} \emph{et~al.}
\newblock \bibinfo{title}{Optical parametric generation in a lithium niobate
  microring with modal phase matching}.
\newblock \emph{\bibinfo{journal}{Physical Review Applied}}
  \textbf{\bibinfo{volume}{11}}, \bibinfo{pages}{034026}
  (\bibinfo{year}{2019}).

\bibitem{bruch201817}
\bibinfo{author}{Bruch, A.~W.} \emph{et~al.}
\newblock \bibinfo{title}{17 000\%/w second-harmonic conversion efficiency in
  single-crystalline aluminum nitride microresonators}.
\newblock \emph{\bibinfo{journal}{Applied Physics Letters}}
  \textbf{\bibinfo{volume}{113}}, \bibinfo{pages}{131102}
  (\bibinfo{year}{2018}).

\bibitem{logan2018400}
\bibinfo{author}{Logan, A.~D.} \emph{et~al.}
\newblock \bibinfo{title}{400\%/w second harmonic conversion efficiency in 14
  $\mu$m-diameter gallium phosphide-on-oxide resonators}.
\newblock \emph{\bibinfo{journal}{Optics express}}
  \textbf{\bibinfo{volume}{26}}, \bibinfo{pages}{33687--33699}
  (\bibinfo{year}{2018}).

\bibitem{petraru2003integrated}
\bibinfo{author}{Petraru, A.}, \bibinfo{author}{Schubert, J.},
  \bibinfo{author}{Schmid, M.}, \bibinfo{author}{Trithaveesak, O.} \&
  \bibinfo{author}{Buchal, C.}
\newblock \bibinfo{title}{Integrated optical mach zehnder modulator based on
  polycrystalline batio 3}.
\newblock \emph{\bibinfo{journal}{Optics letters}}
  \textbf{\bibinfo{volume}{28}}, \bibinfo{pages}{2527--2529}
  (\bibinfo{year}{2003}).

\bibitem{akazawa2005electro}
\bibinfo{author}{Akazawa, H.} \& \bibinfo{author}{Shimada, M.}
\newblock \bibinfo{title}{Electro-optic properties of c-axis oriented linbo3
  films grown on si (1 0 0) substrate}.
\newblock \emph{\bibinfo{journal}{Materials Science and Engineering: B}}
  \textbf{\bibinfo{volume}{120}}, \bibinfo{pages}{50--54}
  (\bibinfo{year}{2005}).

\bibitem{lee2011hybrid}
\bibinfo{author}{Lee, Y.~S.} \emph{et~al.}
\newblock \bibinfo{title}{Hybrid si-linbo 3 microring electro-optically tunable
  resonators for active photonic devices}.
\newblock \emph{\bibinfo{journal}{Optics letters}}
  \textbf{\bibinfo{volume}{36}}, \bibinfo{pages}{1119--1121}
  (\bibinfo{year}{2011}).

\bibitem{chmielak2011pockels}
\bibinfo{author}{Chmielak, B.} \emph{et~al.}
\newblock \bibinfo{title}{Pockels effect based fully integrated, strained
  silicon electro-optic modulator}.
\newblock \emph{\bibinfo{journal}{Optics express}}
  \textbf{\bibinfo{volume}{19}}, \bibinfo{pages}{17212--17219}
  (\bibinfo{year}{2011}).

\bibitem{petraru2001ferroelectic}
\bibinfo{author}{Petraru, A.}, \bibinfo{author}{Siegert, M.},
  \bibinfo{author}{Schmid, M.}, \bibinfo{author}{Schubert, J.} \&
  \bibinfo{author}{Buchal, C.}
\newblock \bibinfo{title}{Ferroelectic batio 3 thin film optical waveguide
  modulators}.
\newblock \emph{\bibinfo{journal}{MRS Online Proceedings Library Archive}}
  \textbf{\bibinfo{volume}{688}} (\bibinfo{year}{2001}).

\bibitem{abel2013strong}
\bibinfo{author}{Abel, S.} \emph{et~al.}
\newblock \bibinfo{title}{A strong electro-optically active lead-free
  ferroelectric integrated on silicon}.
\newblock \emph{\bibinfo{journal}{Nature communications}}
  \textbf{\bibinfo{volume}{4}}, \bibinfo{pages}{1671} (\bibinfo{year}{2013}).

\bibitem{palmer2013low}
\bibinfo{author}{Palmer, R.} \emph{et~al.}
\newblock \bibinfo{title}{Low power mach--zehnder modulator in silicon-organic
  hybrid technology}.
\newblock \emph{\bibinfo{journal}{IEEE Photonics Technology Letters}}
  \textbf{\bibinfo{volume}{25}}, \bibinfo{pages}{1226--1229}
  (\bibinfo{year}{2013}).

\bibitem{xiong2014active}
\bibinfo{author}{Xiong, C.} \emph{et~al.}
\newblock \bibinfo{title}{Active silicon integrated nanophotonics:
  ferroelectric batio3 devices}.
\newblock \emph{\bibinfo{journal}{Nano letters}} \textbf{\bibinfo{volume}{14}},
  \bibinfo{pages}{1419--1425} (\bibinfo{year}{2014}).

\bibitem{jin2014benzocyclobutene}
\bibinfo{author}{Jin, W.} \emph{et~al.}
\newblock \bibinfo{title}{Benzocyclobutene barrier layer for suppressing
  conductance in nonlinear optical devices during electric field poling}.
\newblock \emph{\bibinfo{journal}{Applied Physics Letters}}
  \textbf{\bibinfo{volume}{104}}, \bibinfo{pages}{94\_1}
  (\bibinfo{year}{2014}).

\bibitem{haffner2015all}
\bibinfo{author}{Haffner, C.} \emph{et~al.}
\newblock \bibinfo{title}{All-plasmonic mach--zehnder modulator enabling
  optical high-speed communication at the microscale}.
\newblock \emph{\bibinfo{journal}{Nature Photonics}}
  \textbf{\bibinfo{volume}{9}}, \bibinfo{pages}{525} (\bibinfo{year}{2015}).

\bibitem{abel2016hybrid}
\bibinfo{author}{Abel, S.} \emph{et~al.}
\newblock \bibinfo{title}{A hybrid barium titanate--silicon photonics platform
  for ultraefficient electro-optic tuning}.
\newblock \emph{\bibinfo{journal}{Journal of Lightwave Technology}}
  \textbf{\bibinfo{volume}{34}}, \bibinfo{pages}{1688--1693}
  (\bibinfo{year}{2016}).

\bibitem{heni2017silicon}
\bibinfo{author}{Heni, W.} \emph{et~al.}
\newblock \bibinfo{title}{Silicon--organic and plasmonic--organic hybrid
  photonics}.
\newblock \emph{\bibinfo{journal}{ACS Photonics}} \textbf{\bibinfo{volume}{4}},
  \bibinfo{pages}{1576--1590} (\bibinfo{year}{2017}).

\bibitem{alexander2018nanophotonic}
\bibinfo{author}{Alexander, K.} \emph{et~al.}
\newblock \bibinfo{title}{Nanophotonic pockels modulators on a silicon nitride
  platform}.
\newblock \emph{\bibinfo{journal}{Nature communications}}
  \textbf{\bibinfo{volume}{9}}, \bibinfo{pages}{3444} (\bibinfo{year}{2018}).

\bibitem{abel2019large}
\bibinfo{author}{Abel, S.} \emph{et~al.}
\newblock \bibinfo{title}{Large pockels effect in micro-and nanostructured
  barium titanate integrated on silicon}.
\newblock \emph{\bibinfo{journal}{Nature materials}}
  \textbf{\bibinfo{volume}{18}}, \bibinfo{pages}{42} (\bibinfo{year}{2019}).

\end{thebibliography}


\clearpage
\widetext

\appendix*

\section*{Supplementary Materials to Room-Temperature Photonic Logical Qubits via Second-Order Nonlinearities}

In these supplementary materials we give more details about the literature survey we have performed in order to estimate the feasibility of our design. We also elaborate on the performance of our numerical search techniques.

\subsection{Mapping between the modes of the processor and the modes of the error correcting code}

\begin{figure}[h!]
    \centering
    \includegraphics[width=0.8\textwidth]{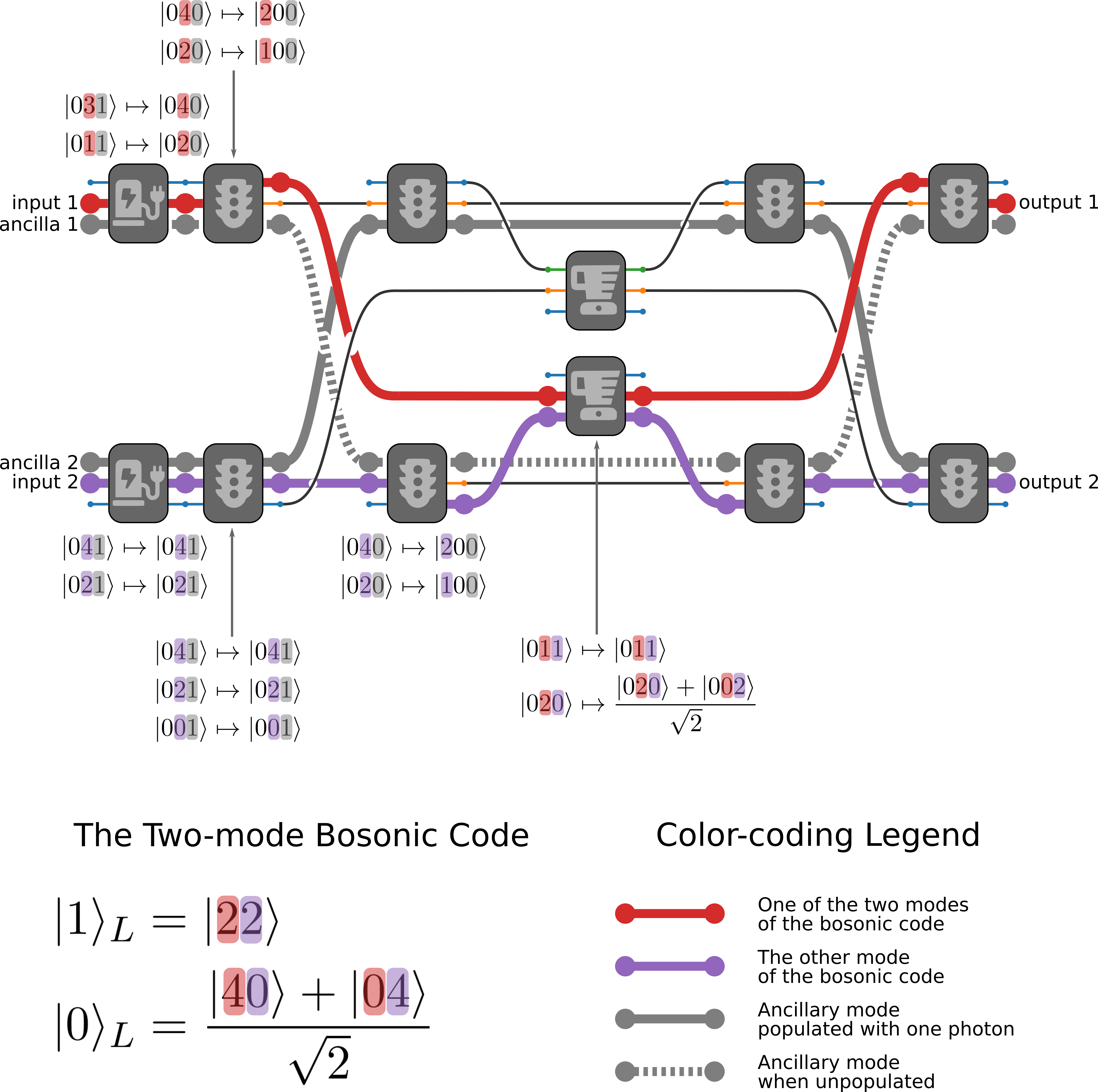}
    \caption{A more detailed rendition of Fig\ \ref{fig:ecc-circuit}, specifically following the trajectory taken by the circuit in the case of a photon loss error on the first of the two modes of the error correcting code (shaded in red). In other words, the input to the circuit is the error state $\alpha |12\rangle + \beta |30\rangle$ and the output is the corrected state $\alpha |22\rangle + \beta \frac{|40\rangle+|04\rangle}{\sqrt 2}$. Gates are annotated with the unitary operations that they perform. The equation insets showing these operations are color coded to make clear the mapping between available hardware modes and code modes.}
    \label{fig:detailed_mapping}
\end{figure}

In Fig.\ \ref{fig:detailed_mapping} we provide a more detailed rendition of how the two modes of the bosonic code we consider are passed through the nonlinear resonators. Notice that a resource of significant importance in this scheme is the "reshuffling" of modes. For instance, the second column of gates routes the code modes (in red and purple) to different spatial modes, depending on whether an ancillary photon was present.

\subsection{Additional examples of multiphoton operations}

For working with single- and dual-rail encodings, we can use linear optics and the already described Toffoli gate. CPhase gates, being similar, but conceptually simpler than Toffoli gates, are also available. When considering higher photon states, our design can be used for additional operations in state preparation, control and measurement. Consider the following application for example:

\paragraph*{Binary Decomposition Gate}

An important advantage of our design is the ease with which it can work on higher-photon-number states. This is crucial if we want to employ bosonic codes or perform number-resolving measurements. Here we demonstrate a gate that transforms a Fock state containing up to four photons of a single mode, into a multi-mode Fock state that contains a binary representation of the initial number of photons. This gate maps $|000\rangle$ and $|001\rangle$ into themselves, and transforms the remaining states as 
\begin{align*}
|002\rangle & \mapsto |011\rangle \\
|003\rangle & \mapsto |101\rangle \\
|004\rangle & \mapsto |111\rangle.
\end{align*}
This gate enables a deterministic photon-number resolving measurement to be made by a set of photodetectors that are not themselves number resolving. This would require only a logarithmic number of detectors, instead of the exponentially large requirements of typical beam splitter trees~\citep{jonsson2019evaluating,young2018general}. Another application is running this gate in reverse in order to prepare interesting higher-photon-number states, e.g., code words of bosonic codes.
Lastly, this gate can be used for heralded generation of high-fidelity Fock states by inputting a weak coherent state and performing post-selection measurements on a subset of the modes.

\subsection{The mapping between the abstract control pulses and the corresponding }

Let us further consider the controllable part of the Hamiltonian. In the main text we have treated $p(t)$ as a dimensionless function. We will now connect it to the quadratures $q(t)$ of the classical field $\mathbf{D}_p(\mathbf{r},t)=q(t)\mathbf{d}_p(\mathbf{r})+\mathrm{c.c}$. Following the same process as above, we derive
\begin{equation}
\begin{split}
\hat{H}_{p}(t) & =
- \frac{\chi^{(2)} \hbar \sqrt{\omega\_b\omega\_c}}{\sqrt{\varepsilon\_0} n^3 \sqrt{V_\text{twm}}}
~ q(t)\hat{b}^\dagger\hat{c} + \mbox{H.c.} \\
\frac{1}{\sqrt{V_{\text{twm}}}} & = \frac{
  \int_{NL} d^{i}_p d^{j*}_b d^{k}_c \mathrm{d}\mathbf{r}}{
  \sqrt{\int |\mathbf{d}_p|^2 \mathrm{d}\mathbf{r} \int |\mathbf{d}_b|^2 \mathrm{d}\mathbf{r} \int |\mathbf{d}_c|^2 \mathrm{d}\mathbf{r}}},
\end{split}
\end{equation}
where $V_{\text{twm}}$ is the mode volume for three-wave mixing. Therefore, the unit of the ordinate in Fig.\ \ref{fig:compiled-gates} is $u_p = \sqrt{\frac{1}{8}\frac{V_\text{tmw}}{V_\text{shg}}\frac{\hbar \omega\_a \omega\_b}{\omega\_c}}$ and the time dependent control field is
\begin{equation}
\begin{split}
   & \mathbf{D}(\mathbf{r}) = p(t) u_p \mathbf{d}_p(\mathbf{r})+\mbox{c.c.} \\
   & \int\frac{|\mathbf{d}_p|^2}{\varepsilon\_0 n^2}\mathrm{d}\mathbf{r}=1,
\end{split}
\end{equation}
where $\mathbf{d}_p$ is the normalized eigenmode of the cavity. In other words, the average number of $\omega\_p$ photons in the control field needs to be just $\frac{1}{4}\frac{V_\text{tmw}}{V_\text{shg}}\frac{\omega\_a \omega\_b}{\omega\_c\omega\_p} |p(t)|^2$. Importantly for design considerations, while a very low $V_\text{shg}$ (i.e., a high overlap between modes $\hat{a}$ and $\hat{b}$) is an indisputable requirement, one can employ a high $V_\text{twm}$ (i.e., low overlap between the control mode at $\omega\_p$ and modes $\hat{b}$ and $\hat{c}$) as long as higher power in the control mode can be tolerated by the material.

\subsection{Literature Survey of Second-harmonic-generation Experiments}

\begin{figure}[h!]
    \centering
    \includegraphics[scale=.8]{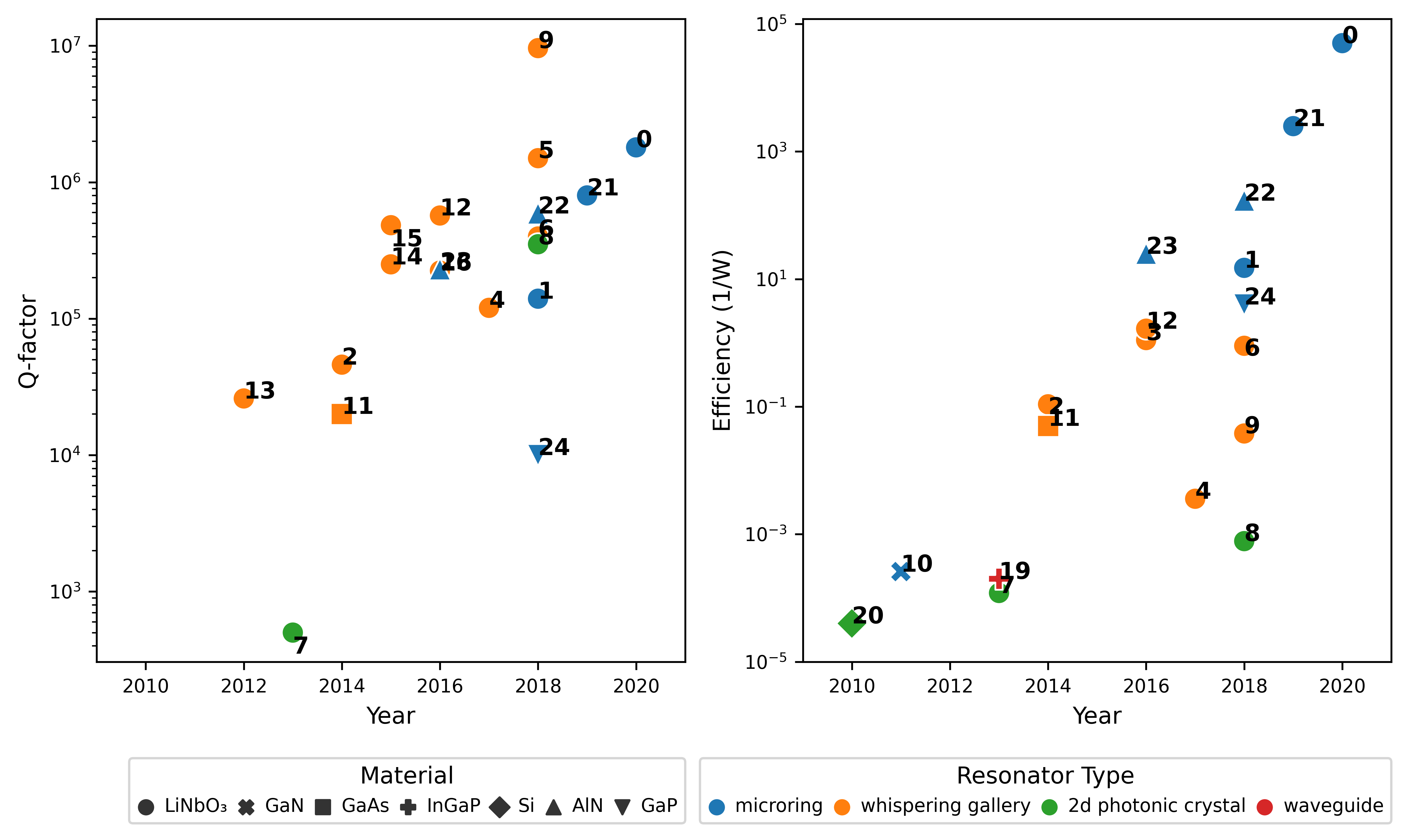}
    \caption{Recent progress in SHG experiments. Numerical labels are linked to the corresponding publication in Table\ \ref{tab:lit_shg_num}.}
    \label{fig:lit_shg_num}
\end{figure}

In Fig.\ \ref{fig:lit_shg_num} we see the progress in second-harmonic-generation (SHG) over the last decade, showing 10 orders of magnitude improvement in measures of efficiency related to our design~\citep{ducci2004continuous,yu2005efficient,galli2010low,xiong2011integrated,wang2012high,diziain2013second,lengle2013efficient,kuo2014second,wang2014integrated,lin2015fabrication,wang2015high,moore2016efficient,lin2016phase,luo2017chip,wang2018high,wolf2018quasi,li2018high,lin2018highly,luo2019optical,lu2019periodically,bruch201817,guo2016second,logan2018400}.
We see the substantial progress in $Q$ factors of photonic mircroresonators.
Lithium niobate is the material of choice due to its high $\chi^{(2)}$. Photonic crystals are promising thanks to their extreme mode confinement, but still need improvements in $Q$ factors. Microrings and whispering gallery resonators hold the majority of high-performance spots for now thanks to a good balance between mode confinement and $Q$ factors. A typical figure of merit in SHG experiments is the efficiency~\citep{guo2016second}  $\eta=\frac{P_\text{out}}{P^2_\text{in}}\propto\frac{Q^3(\chi^{(2)})^2}{V_\text{shg}}$, which is closely related to the number-of-useful-operations figure of merit we are using in the main text. While it is infeasible to directly plot our figure of merit for the results above (due to the vastly diverse hardware in which they were obtained), the exponential growth of the easier-to-measure efficiency bodes well for the future use of our protocol. Note that recently the leading experiments have switched from whispering gallery resonators (orange) to better confined microring resonators (blue), even though their $Q$ factors are much lower, leaving very significant space for further improvements.

\begin{figure}[h!]
    \centering
    \includegraphics[scale=.8]{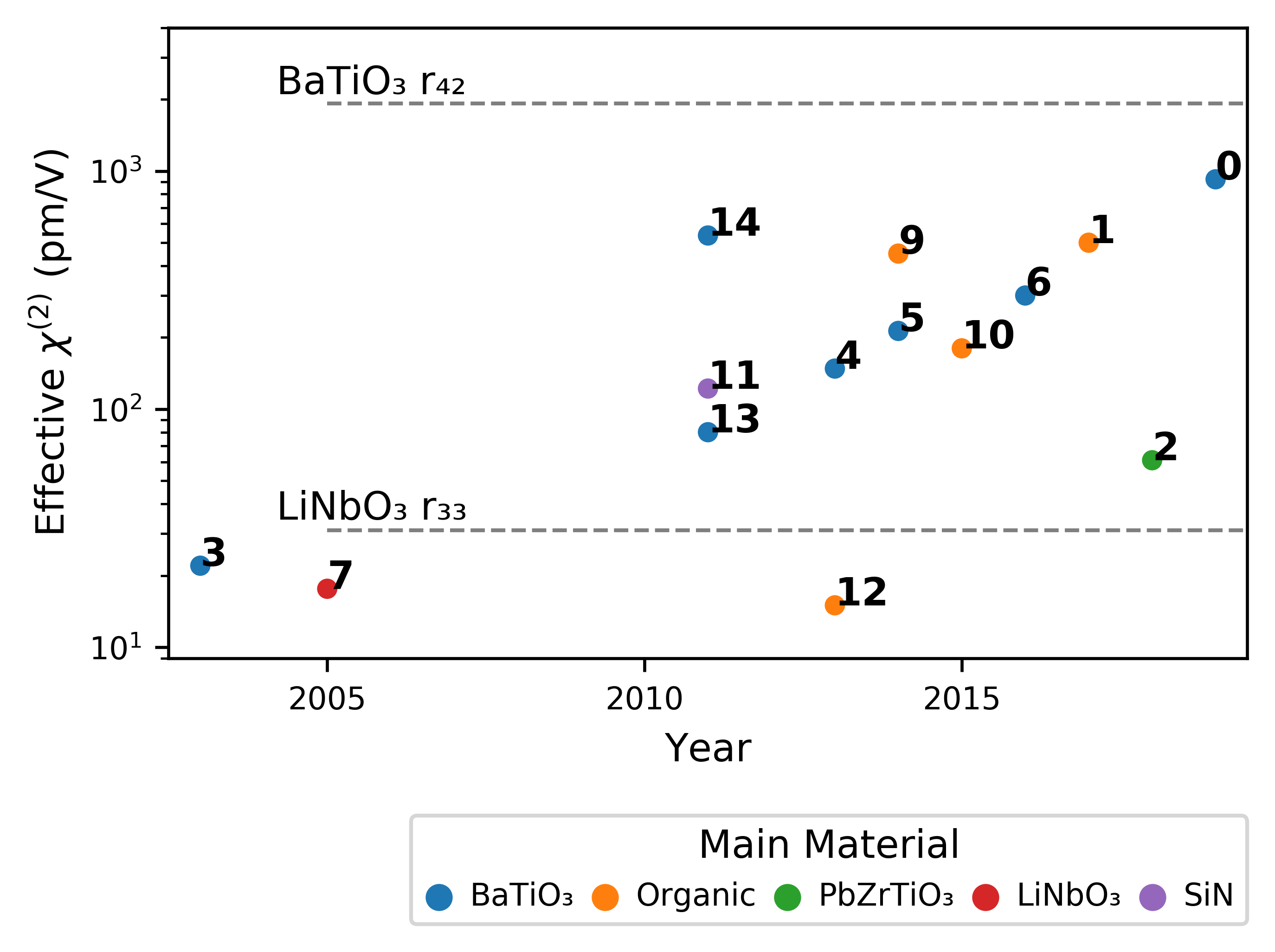}
    \caption{Recent experimental progress in Pockels effect electro-optical modulators. This plot showcases significant process in material science and thin-slab fabrication techniques. Numerical labels are linked to the corresponding publication in Table\ \ref{tab:lit_poc_num}.}
    \label{fig:lit_poc_num}
\end{figure}

Highly nonlinear materials can significantly improve the performance of our protocol. The progress in their development is depicted in Fig.\ \ref{fig:lit_poc_num}. Significant improvements have been achieved both by the discovery of new materials and by placing known materials under strain using novel fabrication techniques. The survey of Pockels-effect electro-optical modulator hardware~\citep{petraru2003integrated,akazawa2005electro,lee2011hybrid,chmielak2011pockels,petraru2001ferroelectic,abel2013strong,palmer2013low,xiong2014active,jin2014benzocyclobutene,haffner2015all,abel2016hybrid,heni2017silicon,alexander2018nanophotonic,abel2019large} shown here reveals the progress in the size of effective electro-optical coefficients for on-chip nonlinear optics. For comparison, the values for these coefficients in bulk crystals are shown in dashed lines. While not all of these results carry over to the optical regime due to frequency dependence, our design can make use of the low-frequency regime if electrical pulses are used to implement the control function $p(t)$. It is particularly interesting that advanced fabrication techniques are capable of inducing record-high $\chi^{(2)}$ values even in materials that do not have second order nonlinearities in their bulk form (e.g., strained SiN in~\citep{chmielak2011pockels}).

\subsection{Optimal Control Pulses}

The main text established the following parameterization for the control pulse used  to construct a given unitary operation $\hat{U}$:

\begin{align}
\hat{U}(\mathrm{\mathbf{v}}) & = 
\prod_{l=1}^{s} \exp\left\{
-i \left[ \operatorname{f}(X_i,P_i)\hat{b}^\dagger\hat{c}
+\sigma(T_i)\hat{a}\hat{b}^{\dagger 2} +\mbox{H.c.}
\right] \right\}  \nonumber  
\end{align} 
where 
\begin{align*} 
 \operatorname{f}(X_i,P_i) & =\arctan(X_i)+i\arctan(P_i) \\
\sigma(T_i) & = \frac{\Delta \tau}{1 + \exp(-T_i)}
\end{align*}
and $\mathbf{v} = \{ X_i, P_i, T_i : i = 1,\ldots, N \}$ is the set of parameters that defines the pulse. 
The parameters $\{ X_i : X_i \in \mathbb{R} \}$ and $\{ P_i : P_i \in \mathbb{R} \}$ are related to the quadrature of the pulse, which is constrained to the interval $[-1,1]$ by $\arctan$, while the $\{ T_i : T_i \in \mathbb{R} \}$ are related to the duration of each segment, which is constrained to the interval $[0,\Delta \tau]$. We fix the number of piecewise-constant intervals, $s$, as well as the relative unitless time scale $\Delta \tau$.

\begin{figure}
    \centering
    \includegraphics[scale=.8]{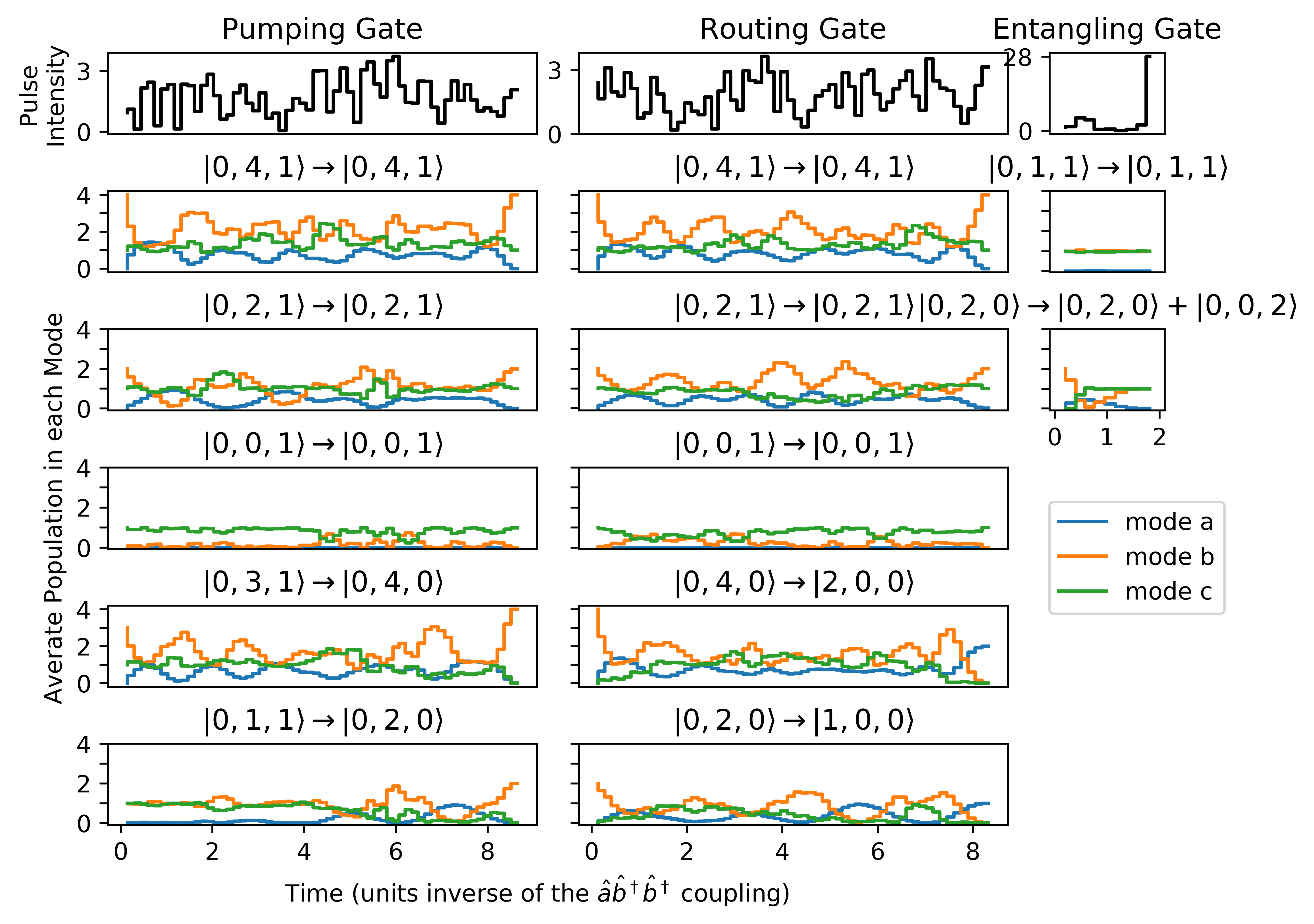}
    \caption{The piece-wise constant control pulses before the smoothing stage. This rich piece-wise constant parameterization admits nearly as much freedom as typical high time-resolution characterizations at much lower computational cost. In this particular case we have used 60 time steps of variable width (instead of the hundreds required for a smooth signal).}
    \label{fig:blocky}
\end{figure}

\begin{figure}
    \centering
    \includegraphics[scale=.8]{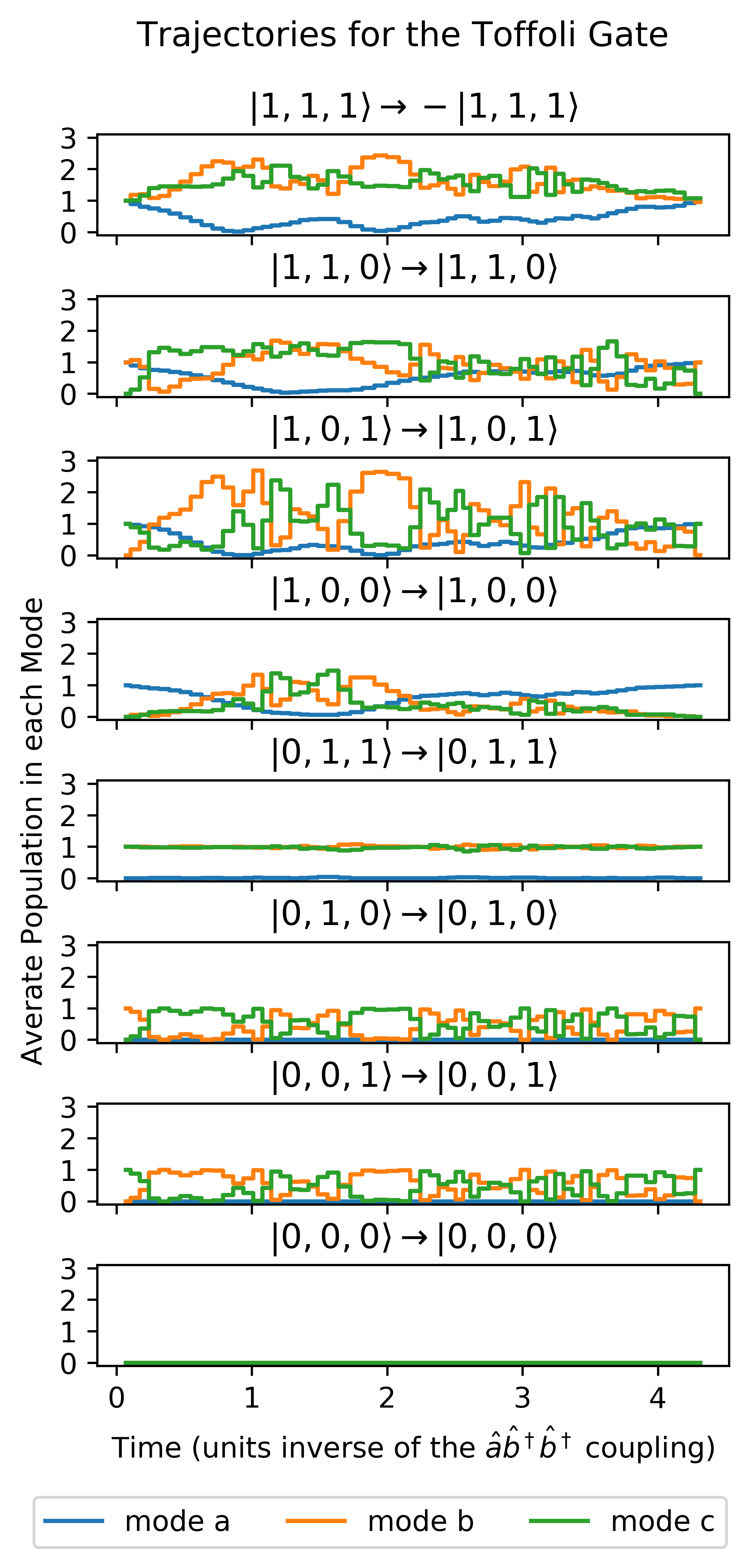}
    \includegraphics[scale=.8]{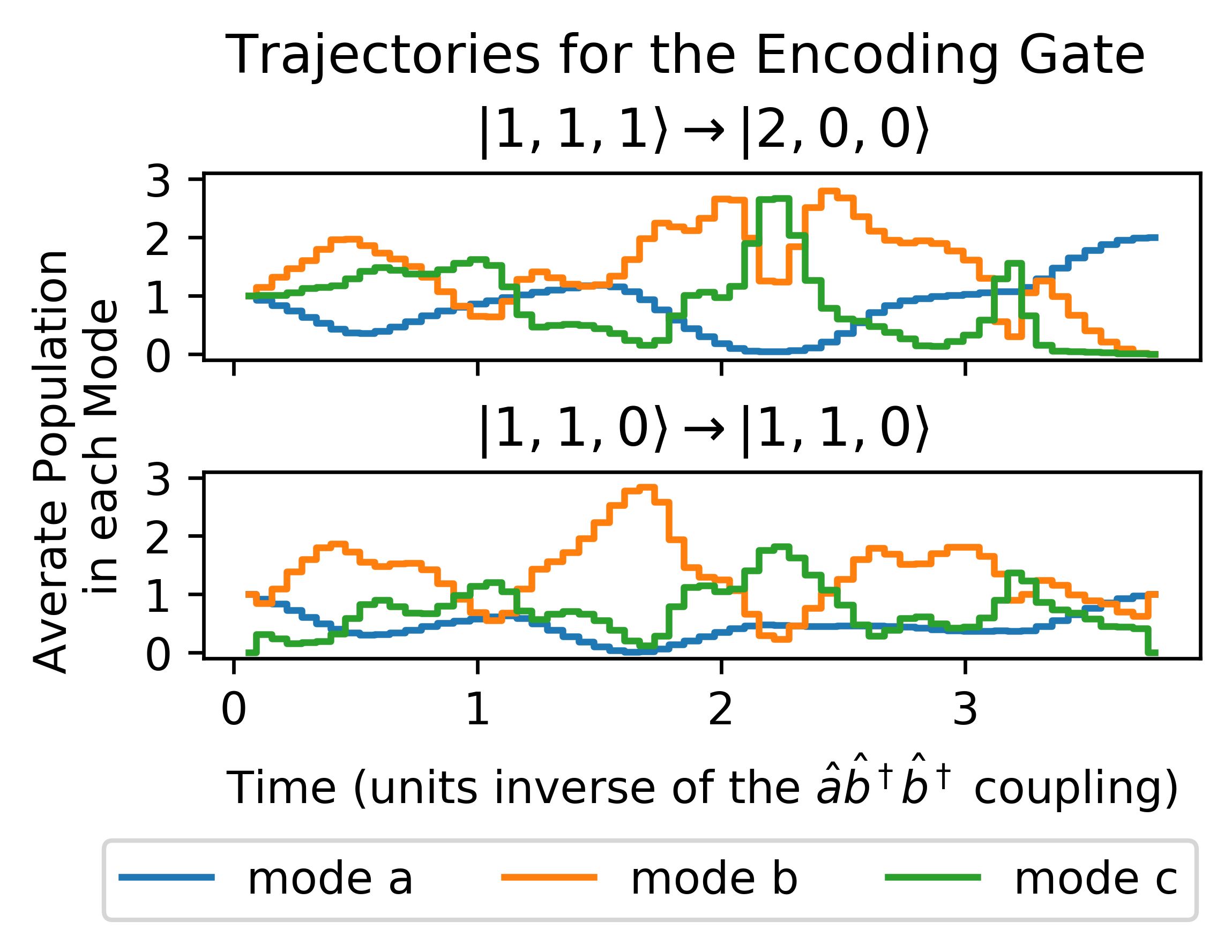}
    \caption{The trajectories undergone by various basis states for the Toffoli gate and the code-encoding gate described in the main text.}
    \label{fig:tof_and_encode}
\end{figure}

As elaborated in the main text, this produces a well constrained control pulse with a small number of piecewise-constant intervals (for instance, we used 60 steps). This leads to faster convergence and it is numerically much less taxing than other approaches. However, as a second step we use typical optimal control techniques (like GRAPE) with much higher time-resolution in order to smooth the control pulses. The figures in the main text show this smoothed control pulse, but here in Fig.\ \ref{fig:blocky} and Fig.\ \ref{fig:tof_and_encode} we show what the initial piecewise-constant pulses look like. In particular, Fig.\ \ref{fig:tof_and_encode} also shows the trajectories for the Toffoli and code-encoding gates, which were described only in prose in the main text.

\begin{figure}
    \centering
    \includegraphics[scale=.8]{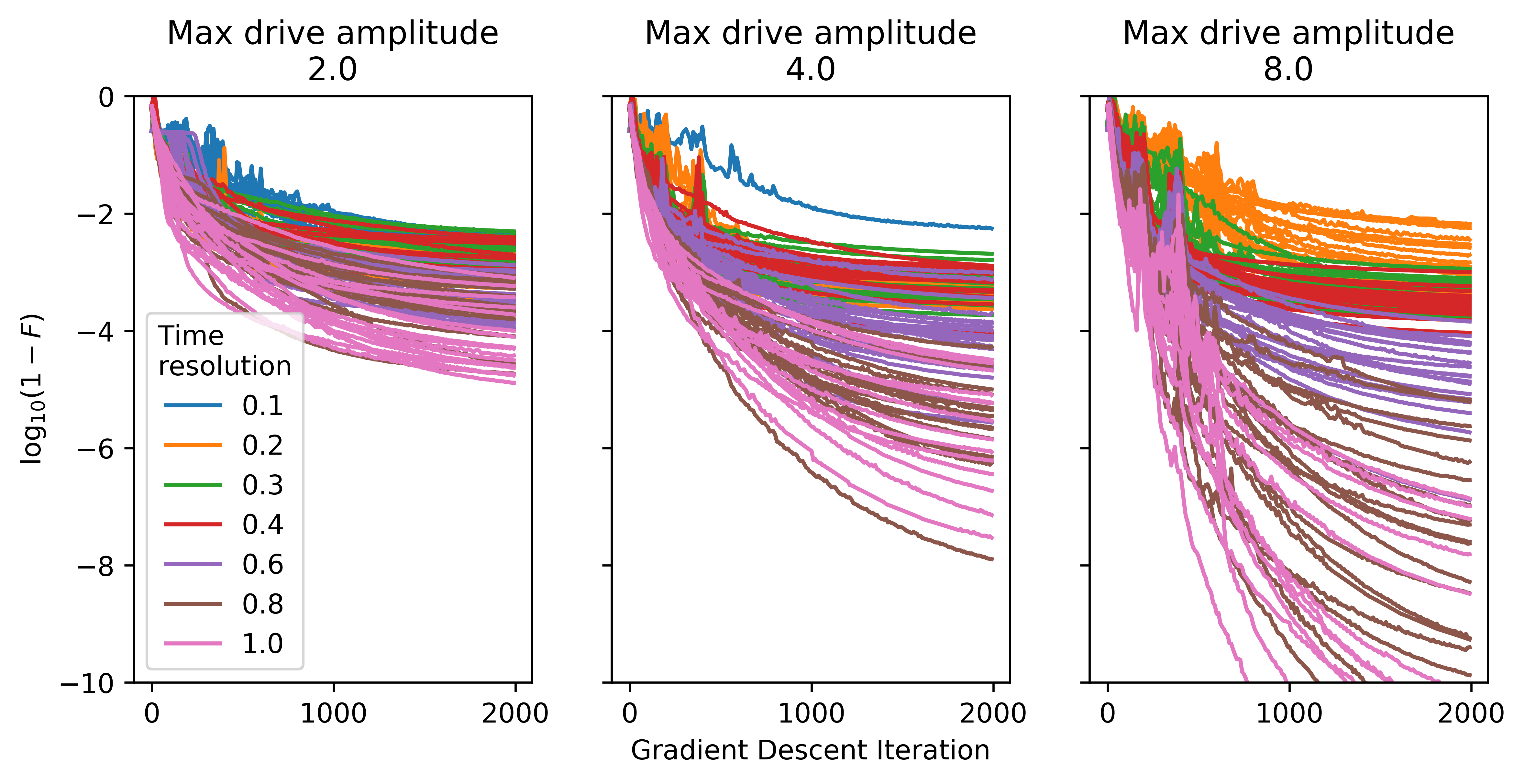}
    \caption{The gradient descent optimization procedure performs differently depending on the constraints we have put on our control pulses. The plots above show the infidelity $1-F$ for a control pulse implementing a particular unitary operation as a function of the iteration steps of the gradient descent optimizer. Each facet depicts different unitless maximum amplitude being permitted for the control drive. Colors represent the unitless characteristic duration $\Delta \tau$ for each piece-wise constant piece of the control drive. We use 60 such pieces in each of the control pulses being considered. There are multiple lines of the same color as we rerun the optimization with varying initial guesses for the control pulse. We see that the gradient descent consistently finds a solution, but plateaus to rather high infidelities of 0.01 if the pulse is overconstrained (e.g. too short or too weak). As we permit higher amplitudes and longer pulses we see that the gradient descent finds much better pulses with infidelities approaching the floating point numerical floor.}
    \label{fig:gradient_iterations}
\end{figure}

\begin{figure}
    \centering
    \includegraphics[scale=.8]{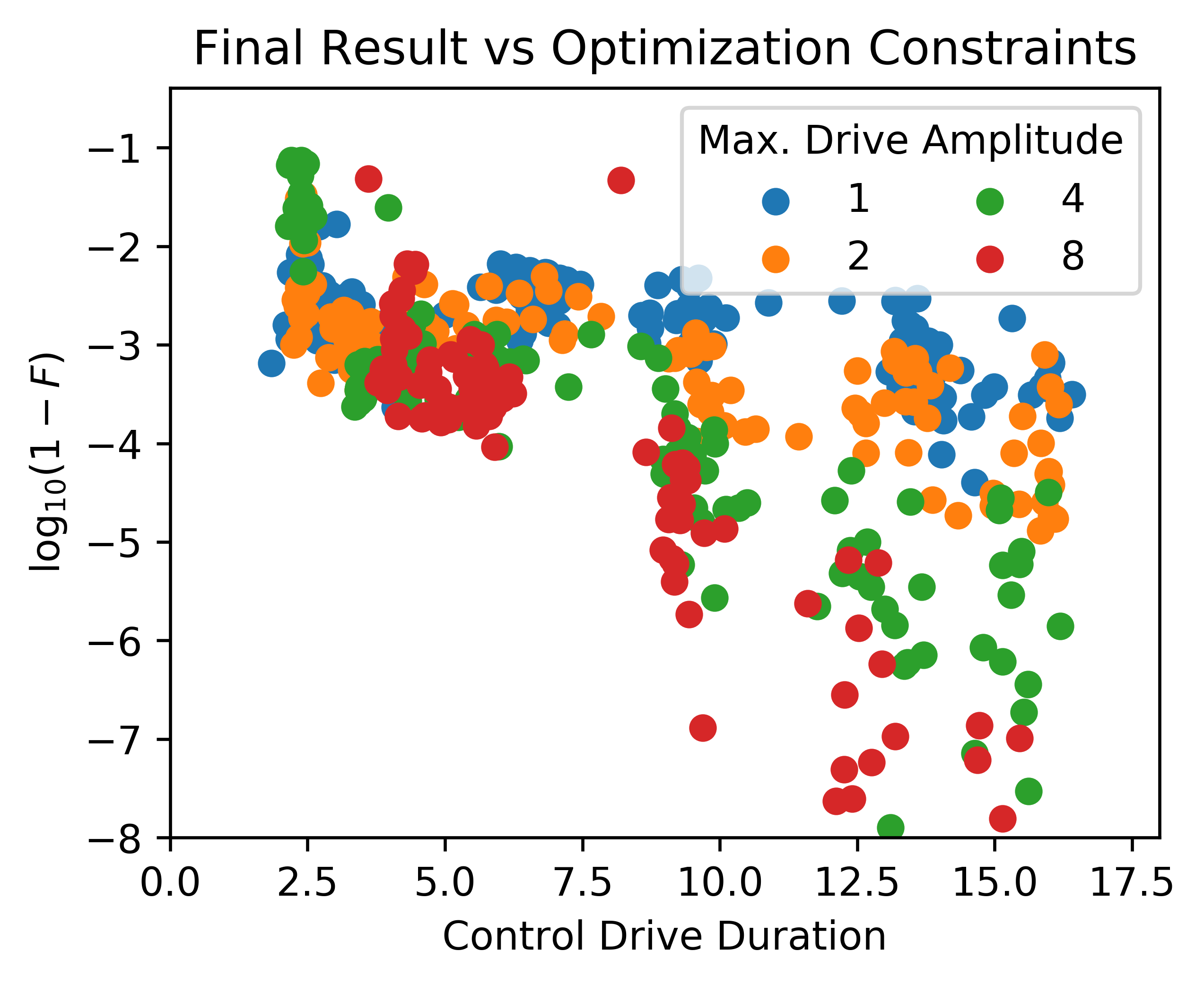}
    \caption{This plot shows the same data as Fig.\ \ref{fig:gradient_iterations}, focusing on the performance of the designed unitary operation after the gradient descent has concluded (i.e. only the points at the 2000th iteration from Fig.\ \ref{fig:gradient_iterations}). We again see that less constrained pulses (higher permitted amplitude or duration) permit much higher fidelities, approaching the floating point numerical limits. In particular, the infidelity drops exponentially with the length of the pulse. This would be countered with the exponential growth of the chance for photon loss as the pulse duration increases, leading to an optimal pulse duration compromising between the two effects.}
    \label{fig:fidelity_vs_duration}
\end{figure}

Lastly, Fig.\ \ref{fig:gradient_iterations} and Fig.\ \ref{fig:fidelity_vs_duration} showcase the good performance of our parameterization under gradient descent. In particular, we see that as we relax the various regularization constraints (like permitting higher amplitudes or longer durations) we obtain better performing control pulses.

\pagebreak

\begin{table}[h!]
    \centering
\begin{tabular}{lll}
{} &  Year &                                                                                                                   Publication \\
0 &  2020 & \href{https://arxiv.org/abs/2007.07411}{Towards 1\% single photon nonlinearity with periodically-poled lithium niobate microring resonators} \\
1 &  2018 & \href{https://journals.aps.org/prapplied/abstract/10.1103/PhysRevApplied.11.034026}{Optical Parametric Generation in a Lithium Niobate Microring with Modal Phase Matching} \\
2 &  2014 & \href{https://www.osapublishing.org/oe/abstract.cfm?uri=oe-22-25-30924}{Integrated high quality factor lithium niobate microdisk resonators} \\
3 &  2016 & \href{https://journals.aps.org/prapplied/abstract/10.1103/PhysRevApplied.6.014002}{Phase-Matched Second-Harmonic Generation in an On-Chip LiNbO3 Microresonator} \\
4 &  2017 & \href{https://www.osapublishing.org/oe/abstract.cfm?uri=oe-25-20-24531}{On-chip second-harmonic generation and broadband parametric down-conversion in a lithium niobate microresonator } \\
5 &  2018 & \href{https://www.osapublishing.org/ol/abstract.cfm?uri=ol-43-12-2917}{High-Q chaotic lithium niobate microdisk cavity} \\
6 &  2018 & \href{https://www.osapublishing.org/optica/abstract.cfm?uri=optica-5-7-872}{Quasi-phase-matched nonlinear optical frequency conversion in on-chip whispering galleries} \\
7 &  2013 & \href{https://aip.scitation.org/doi/abs/10.1063/1.4817507}{Second harmonic generation in free-standing lithium niobate photonic crystal L3 cavity} \\
8 &  2018 & \href{https://arxiv.org/abs/1806.04755}{High-Q two-dimensional lithium niobate photonic crystal slab nanoresonators} \\
9 &  2018 & \href{https://arxiv.org/abs/1809.04523}{Highly-efficient second and third harmonic generation in a monocrystalline lithium niobate microresonator} \\
10 &  2011 & \href{https://www.osapublishing.org/oe/abstract.cfm?uri=oe-19-11-10462}{Integrated GaN photonic circuits on silicon (100) for second harmonic generation } \\
11 &  2014 & \href{https://www.nature.com/articles/ncomms4109}{Second-harmonic generation using -quasi-phasematching in a GaAs whispering-gallery-mode microcavity} \\
12 &  2016 & \href{https://ieeexplore.ieee.org/abstract/document/7788916}{Efficient second harmonic generation in lithium niobate on insulator} \\
13 &  2012 & \href{https://www.osapublishing.org/oe/abstract.cfm?uri=oe-20-27-28119}{High-quality LiNbO3 microdisk resonators by undercut etching and surface tension reshaping} \\
14 &  2015 & \href{https://www.nature.com/articles/srep08072}{Fabrication of high-Q lithium niobate microresonators using femtosecond laser micromachining} \\
15 &  2015 & \href{https://www.osapublishing.org/oe/abstract.cfm?uri=oe-23-18-23072}{High-Q lithium niobate microdisk resonators on a chip for efficient electro-optic modulation } \\
16 &  2016 & \href{https://www.nature.com/articles/srep36920}{Chip-scale cavity optomechanics in lithium niobate} \\
17 &  2004 & \href{https://aip.scitation.org/doi/abs/10.1063/1.1703847}{Continuous-wave second-harmonic generation in modal phase matched semiconductor waveguides } \\
18 &  2005 & \href{https://www.osapublishing.org/oe/abstract.cfm?uri=oe-13-26-10742}{Efficient continuous wave second harmonic generation pumped at 1.55 \mbox{$\mu$}m in quasi-phase-matched AlGaAs waveguides } \\
19 &  2013 & \href{https://aip.scitation.org/doi/abs/10.1063/1.4802790}{Efficient second harmonic generation in nanophotonic waveguides for optical signal processing} \\
20 &  2010 & \href{https://www.osapublishing.org/oe/abstract.cfm?uri=oe-18-25-26613}{Low-power continuous-wave generation of visible harmonics in silicon photonic crystal nanocavities} \\
21 &  2019 & \href{https://www.osapublishing.org/optica/abstract.cfm?uri=optica-6-12-1455}{Periodically poled thin-film lithium niobate microring resonators with a second-harmonic generation efficiency of 250,000\%/W} \\
22 &  2018 & \href{https://aip.scitation.org/doi/abs/10.1063/1.5042506}{17000\%/W second-harmonic conversion efficiency in single-crystalline aluminum nitride microresonators} \\
23 &  2016 & \href{https://www.osapublishing.org/optica/abstract.cfm?uri=optica-3-10-1126}{Second-harmonic generation in aluminum nitride microrings with 2500\%/W conversion efficiency} \\
24 &  2018 & \href{https://www.osapublishing.org/oe/abstract.cfm?uri=OE-26-26-33687}{400\%/W second harmonic conversion efficiency in 14 $\mu$m-diameter gallium phosphide-on-oxide resonators }
\end{tabular}
    \caption{Table of recent SHG experiments featured in Fig.\ \ref{fig:lit_shg_num}. Each row is hyperlinked to the corresponding publication.}
    \label{tab:lit_shg_num}
\end{table}

\begin{table}[h!]
    \centering
\begin{tabular}{lll}
{} &  Year &                                                                                                                   Publication \\
0 &  2019 & \href{https://www.nature.com/articles/s41563-018-0208-0}{Large Pockels effect in micro- and nanostructured barium titanate integrated on silicon} \\
1 &  2017 & \href{https://pubs.acs.org/doi/abs/10.1021/acsphotonics.7b00224}{Silicon–Organic and Plasmonic–Organic Hybrid Photonics} \\
2 &  2018 & \href{https://www.nature.com/articles/s41467-018-05846-6}{Nanophotonic Pockels modulators on a silicon nitride platform} \\
3 &  2003 & \href{https://www.osapublishing.org/ol/abstract.cfm?uri=ol-28-24-2527}{Integrated optical Mach–Zehnder modulator based on polycrystalline BaTiO3} \\
4 &  2013 & \href{https://www.nature.com/articles/ncomms2695}{A strong electro-optically active lead-free ferroelectric integrated on silicon} \\
5 &  2014 & \href{https://pubs.acs.org/doi/abs/10.1021/nl404513p}{Active Silicon Integrated Nanophotonics: Ferroelectric BaTiO3 Devices} \\
6 &  2016 & \href{https://www.osapublishing.org/jlt/abstract.cfm?uri=jlt-34-8-1688}{A Hybrid Barium Titanate–Silicon Photonics Platform for Ultraefficient Electro-Optic Tuning} \\
7 &  2005 & \href{https://www.sciencedirect.com/science/article/pii/S0921510705000784}{Electro-optic properties of c-axis oriented LiNbO3 films grown on Si(1 0 0) substrate} \\
8 &  2011 & \href{https://www.osapublishing.org/ol/abstract.cfm?uri=ol-36-7-1119}{Hybrid Si-LiNbO${}_3$ microring electro-optically tunable resonators for active photonic devices } \\
9 &  2014 & \href{https://aip.scitation.org/doi/full/10.1063/1.4884829}{Benzocyclobutene barrier layer for suppressing conductance in nonlinear optical devices during electric field poling} \\
10 &  2015 & \href{https://www.nature.com/articles/nphoton.2015.127}{All-plasmonic Mach–Zehnder modulator enabling optical high-speed communication at the microscale} \\
11 &  2011 & \href{https://www.osapublishing.org/oe/abstract.cfm?uri=oe-19-18-17212}{Pockels effect based fully integrated, strained silicon electro-optic modulator} \\
12 &  2013 & \href{https://ieeexplore.ieee.org/abstract/document/6510430}{Low Power Mach–Zehnder Modulator in Silicon-Organic Hybrid Technology} \\
13 &  2011 & \href{https://www.cambridge.org/core/journals/mrs-online-proceedings-library-archive/article/ferroelectic-batio3-thin-film-optical-waveguide-modulators/659073590F226186294B8D23C0700E64}{Ferroelectic BaTiO3 Thin Film Optical Waveguide Modulators} \\
14 &  2011 & \href{https://www.cambridge.org/core/journals/mrs-online-proceedings-library-archive/article/ferroelectic-batio3-thin-film-optical-waveguide-modulators/659073590F226186294B8D23C0700E64}{Ferroelectic BaTiO3 Thin Film Optical Waveguide Modulators} \\
\end{tabular}
    \caption{Table of recent electro-optical modulator experiments featured in Fig.\ \ref{fig:lit_poc_num}. Each row is hyperlinked to the corresponding publication.}
    \label{tab:lit_poc_num}
\end{table}

\end{document}